\documentclass[12pt]{iopart}
\usepackage{iopams}

\usepackage{amssymb}
\usepackage{amsfonts}
\usepackage{bbold}
\usepackage{graphicx}
\usepackage{epsfig}
\usepackage{color}
\usepackage{url}
\usepackage{times}
\usepackage{bm}
\usepackage{mathrsfs}
\usepackage[utf8]{inputenc}
\usepackage{hyperref}
\usepackage{enumerate}
\usepackage{amsthm}
\usepackage{verbatim}
\usepackage{cite}
\usepackage{bbm}
\usepackage{stmaryrd}
\usepackage{slashed}
\usepackage{tikz}
\usepackage{caption}
\usepackage{subcaption}
\usepackage{revsymb} %needed to define \lambdabar
\usepackage{subeqn}

\newcommand{\beq}{\begin{equation}}
\newcommand{\eeq}{\end{equation}}
\newcommand{\bea}{\begin{eqnarray}}
\newcommand{\eea}{\end{eqnarray}}
\newcommand{\bit}{\begin{itemize}}
\newcommand{\eit}{\end{itemize}}
\newcommand{\ben}{\begin{enumerate}}
\newcommand{\een}{\end{enumerate}}
\newcommand{\nn}{\nonumber}
\newcommand{\dfrac}[2]{{\displaystyle\frac{#1}{#2}}}
\newcommand{\eqref}[1]{(\ref{#1})}

\newcommand{\flatten}[1]{{\,\rm flatten}\left[ #1 \right]}

\newcommand{\rh}{r_{h}}
\newcommand{\rc}{r_{c}}

\newcommand{\spin}{\frak s}
\newcommand{\BLCoord}{t, r, \theta, \phi}
\newcommand{\spinPsi}[1]{{}_{#1}\!\Psi}
\newcommand{\spinPhio}[1]{{}_{#1}\!\Phi_{m;\omega}(r,\theta)}

\newcommand{\spinRo}[1]{{}_{#1}R_{m;\omega}(r)}
\newcommand{\spinSo}[1]{{}_{#1}S_{m;\omega}(\theta)}

\newcommand{\spinS}[1]{{}_{#1}S_{\ell m;\omega}(\theta)}
\newcommand{\spinR}[1]{{}_{#1}R_{n \ell m}(r)}

\newcommand{\spinY}[1]{{}_{#1}Y_{\ell m;\omega}(\theta, \phi)}
\newcommand{\spinlambda}[2]{{}_{#1}\lambdabar_{\ell m;#2}}
\newcommand{\spinlambdaQNM}[1]{{}_{#1}\lambdabar_{n \ell m}}

\def\scri{\mathscr{I}}
\def\H{\mathscr{H}}

\newcommand{\HypCoord}{\tau, \sigma, x, \varphi}
\newcommand{\spinHyperPsi}[1]{{}_{#1}\!\bar\Psi}

\newcommand{\spinHypRo}[1]{{}_{#1}\bar R_{m;\bar \omega}(\sigma)}
\newcommand{\spinHypSo}[1]{{}_{#1}\bar S_{m;\bar \omega}(x)}

\newcommand{\spinHyperPhi}[1]{{}_{#1}\!\bar\Phi_{m;\bar \omega} (\sigma, x)}
\newcommand{\spinHyperPhio}[1]{{}_{#1}\!\bar\Phi_{m;\bar \omega}}

\newcommand{\spinHyperPhioQNM}[1]{{}_{#1}\!\bar\Phi_{m;\bar \omega_{q,\ell,m}}}

\newcommand{\spinHyperPhiQNM}[4]{{}_{#1}\!\bar\Phi_{#2,#3,#4}}

\newcommand{\vecspinHyperPhi}[1]{{}_{#1}\!\vec{\boldsymbol{\bar\Phi}}_{m;\bar \omega}}

\newcommand{\spinHyperUpsilono}[1]{{}_{#1}\!\bar\Upsilon_{m;\bar \omega}}

\newcommand{\NBI}{\address{Center of Gravity, Niels Bohr Institute, Blegdamsvej 17, 2100 Copenhagen, Denmark}}
\newcommand{\SL}{\address{Ecole Normale Supérieure de Lyon, 46 All. d'Italie, 69007 Lyon, France}}

\begin{document}

\title{Quasinormal modes in Kerr spacetime as a 2D Eigenvalue problem}
\author{Jamil Assaad}
\SL
%\NBI

\author{Rodrigo Panosso Macedo}
\NBI
\date{June 2025}

\begin{abstract}
We revisit the computation of quasinormal modes (QNMs) of the Kerr black hole using a numerical approach exploiting a representation of the Teukolsky equation as a $2D$ elliptic partial differential equation. By combining the hyperboloidal framework with a $m$-mode decomposition, we recast the QNM problem into a genuine eigenvalue problem for each azimuthal mode. This formulation enables the simultaneous extraction of multiple QNMs, traditionally labelled by overtone number $n$ and angular index $\ell$, without requiring prior assumptions about their structure. We advocate for a simplified notation in which each overtone is uniquely labelled by a single index $q$, thereby avoiding the conventional but artificial distinction between regular and mirror modes. We compare two distinct hyperboloidal gauges—radial fixing and Cauchy horizon fixing—and demonstrate that, despite their different geometric properties and behaviour in the extremal limit, they yield numerical values for the QNM spectra with comparable accuracy and exponential convergence. Moreover, we show that strong gradients observed near the horizon in the extremal Kerr regime are coordinate artefacts of specific slicing rather than physical features. Finally, we investigate the angular structure of the QNM eigenfunctions and show that the $m$-mode approach allows flexible projection onto both spin-weighted spheroidal and spherical harmonic bases. These results underscore the robustness and versatility of the hyperboloidal $m$-mode method as a foundation for future studies of QNM stability, pseudospectra, and mode excitation in gravitational wave astronomy.
\end{abstract}

\maketitle

\setlength{\parindent}{0pt}

\section{Introduction}
The Teukolsky equation (TE) is a cornerstone of black hole perturbation theory \cite{TeukolskyPRL,Teukolsky1,Teukolsky2,Teukolsky3,Chandrasekhar:1985kt}. By describing the dynamics of linearized curvature perturbations, electromagnetic fields, or scalar fields on the Kerr background, it plays a crucial role in formal approaches to proving the stability of this spacetime \cite{Whiting:1988vc,Dafermos:2017yrz,Andersson:2019dwi}, as well as in the wave modeling for gravitational wave astronomy \cite{Pound:2021qin,LISAConsortiumWaveformWorkingGroup:2023arg,Berti:2025hly}. In particular, the notion of quasi-normal modes (QNMs) \cite{Kokkotas,Nollert,BertiCardoso,Konoplya:Zhidenko} takes center stage in theoretical studies of solutions to the TE and the so-called black hole spectroscopy program \cite{Dreyer:2003bv,BertiCardoso2,Berti:2025hly}. It is therefore unsurprising that there is a continuous effort to improve methods for solving the TE in various contexts and for calculating key properties of its solutions, such as QNMs\cite{Lo:2025njp,Berti:2025hly}.

\smallskip

A remarkable property of the TE is its separability into a set of ordinary differential equations (ODEs) when expressed in the frequency domain \cite{Teukolsky1}. Broadly speaking, the TE constitutes a set of decoupled wave-like equations for master functions $\spinPsi{\spin}$, with $\spin$ denoting the function’s spin weight. Considering a parametrization of the Kerr spacetime in terms of spherical-like coordinates $(\BLCoord)$—traditionally the Boyer–Lindquist coordinates—the existence of the Killing vectors $\partial_t$ and $\partial_\phi$, associated respectively with time translations and rotations around the black hole’s symmetry axis, allows us to apply a Fourier (or Laplace) transformation to map the time and angular dependence into parameters $(m, \omega)$ via $\spinPsi{\spin}(\BLCoord) \sim e^{-i \omega t} e^{i m \phi} \spinPhio{\spin}$. Then, an ansatz of the form $\spinPhio{\spin} \sim \spinRo{\spin} \spinSo{\spin}$ yields independent ODEs for each function $\spinRo{\spin}$ and $\spinSo{\spin}$, related to each other via a separation constant $\lambdabar$.

\smallskip
The solutions to the angular sector constitute the so-called spin-weighted spheroidal harmonics $\spinY{\spin} \sim \spinS{\spin} e^{i m \phi}$, where the index $\ell \in \mathbb{N}$ labels the discrete set of regular solutions $\spinS{\spin}$, understood as eigenfunctions of the angular operator with an associated eigenvalue $\spinlambda{\spin}{\omega}$.
A second eigenvalue problem arises in the radial direction when the radial ODE is supplemented with boundary conditions corresponding to the physically relevant scenario. The condition that the field’s energy propagates toward the wave zone at $r \rightarrow \infty$ or is absorbed by the black hole event horizon $\rh$ yields discrete set of eigenfunctions $\spinR{\spin}$ associated with eigenvalues $\omega_{n\ell m}$, i.e. the QNMs. Note that, despite the separability of the angular and radial equations, the QNM eigenvalue problem forms a system of two ODEs coupled by the frequency parameter $\omega_{n\ell m}$ and the separation constant $\spinlambdaQNM{\spin} := \spinlambda{\spin}{\omega_{n\ell m}}$.

\smallskip
There are several approaches to solving the angular and radial ODEs. Analytically, both equations can be expressed in terms of the so-called confluent Heun equation \cite{Marcilhacy1983,Blandin1983,Hortacsu:2011rr,Minucci:2024qrn}. Numerically, Leaver’s method remains one of the most efficient strategies for solving the coupled system of radial and angular equations \cite{Leaver,Nollert:1993zz,Cook:2014cta}. In a nutshell, the method explicitly incorporates the asymptotic boundary behaviour of the QNM eigenfunctions as a pre-factor and expands the remaining regular part into a Taylor series around the event horizon. The Taylor coefficients then satisfy a recurrence relation parametrized by the frequency $\omega$ and separation constant $\lambdabar$. The QNM frequencies $\omega_{n \ell m}$ are the complex values that yield a minimal solution to the recurrence relation. However, the radial equation cannot be solved independently of the angular one. The coupling between the angular and radial equations via the QNM frequency and separation constant requires a root-finding algorithm to solve both ODEs together, having the pair $(\omega_{n\ell m},\spinlambdaQNM{\spin})$ as unknown variables.

\smallskip
Despite the success of this approach \cite{Cook:2014cta}, some fundamental drawbacks remain in the traditional formulation of the QNM eigenvalue problem. The most notable is the blowup of the radial functions $\spinR{\spin}$ as $r\rightarrow \infty$ or $r \rightarrow \rh$. It is now well established \cite{Zenginoglu:2007jw,ZenginogluGeoBH,PanossoSphSym,PanossoMacedo:2024nkw} that such ill behaviour arises from a poor choice of the spacetime coordinate system, where hypersurfaces of constant time $t$ accumulate at the bifurcation sphere ${\cal B}$ and future infinity $i^0$. Adapting the time foliation to the causal structure of the black hole event horizon $\H^+$ and future null infinity $\scri^+$ via a hyperboloidal coordinate system resolves this issue. Moreover, the hyperboloidal framework opens new possibilities in black hole perturbation theory by incorporating tools from the theory of non-normal operators \cite{jaramillo2021pseudospectrum,Jaramillo:2022kuv}.

\smallskip
The theoretical and numerical infrastructure for hyperboloidal frameworks in spherically symmetric spacetimes has seen significant progress in the past decades \cite{PanossoSphSym}. For the Kerr spacetime, most efforts have been focused on solving the Teukolsky equation in the time domain \cite{Racz:2011qu,Jasiulek:2011ce,Harms:2014dqa,Panosso:Ansorg,Csukas:2019kcb,Csukas:2021sia,Markakis:2023pfh} or on considering constraint equations for Kerr-like data serving as initial data for eventual nonlinear numerical evolutions \cite{Schinkel,Schinkel2}. Recently, however, different hyperboloidal strategies for the Kerr spacetime have been systematically unified into a single formalism \cite{PanossoKERR}, where the notion of the hyperboloidal minimal gauge plays a crucial role. By retaining only the minimal conditions necessary for constructing a hyperboloidal foliation, this approach introduces a relatively simple family of hyperboloidal coordinates for the Kerr spacetime and, building on the foundational work~\cite{ZenginogluGeoBH}, it establishes the main treatment of the problem in the frequency domain. Indeed, Ref.~\cite{PanossoKERR} demonstrates that Leaver’s strategy is precisely the frequency-domain treatment of a hyperboloidal Teukolsky wave equation.

\smallskip
In this context, Ref.~\cite{Ripley:2022ypi} developed a QNM solver for the TE based on the hyperboloidal framework. The strategy follows the same steps as previously described: one exploits the separability of the TE in the frequency domain and solves the eigenvalue problem for the quantities $(\omega_{n\ell m},\spinlambdaQNM{\spin})$ resulting from the radial and angular equations. The crucial difference is that the radial equation can now be solved directly, without requiring any form of regularization scheme. Thus, a spectral method based on Chebyshev polynomials\cite{jaramillo2021pseudospectrum} (or adaptation thereof \cite{Zhu:2023mzv}) is a useful numerical strategy to tackle the eigenvalue problem. In particular, Ref.~\cite{Ripley:2022ypi} raises the necessity of further studying the QNM eigenfunctions in the near-extremal Kerr limit as the author finds functions with strong gradient near the horizon, which might be related to the Aretakis instability\cite{Aretakis:2011gz,Aretakis:2012ei}.

\smallskip
Unfortunately, an approach exploiting the equations's separability properties has a drawback compared to its counterpart in spherical symmetry. In spherical symmetry, the radial and angular equations completely decouple. Since the angular equation becomes independent of the frequency $\omega_{n\ell m}$, the angular eigenvalue problem is solved analytically for the separation constant, yielding $\spinlambdaQNM{\spin} = \spin(\spin +1) - \ell (\ell+1)$. Consequently, the numerical scheme treating the the radial equation as a genuine eigenvalue problem determines {\em simultaneously}~\cite{jaramillo2021pseudospectrum} all QNM overtones $n=0, 1, \cdots$. In other words, solving the equations as a genuine eigenvalue problem eliminates the need for an initial guess of a particular $\omega_{n\ell m}$ value in a root-finding algorithm. However, in the axisymmetric configuration, the double eigenvalue problem still requires a root-finding scheme, where an initial guess for $(\omega_{n\ell m},\spinlambdaQNM{\spin})$ is necessary. In fact, one can also represent the radial function in terms of a Taylor series rather than a Chebyshev approximation, yielding a methodology identical to Leaver’s approach --- perhaps just formulated in a different hyperboloidal gauge.

\smallskip
The current work explores an alternative strategy that completely bypasses the separation constant $\spinlambdaQNM{\spin}$ by formulating the frequency-domain TE in an $m$-mode scheme. By working with a 2D elliptic partial differential equation (PDE), the QNM problem is directly cast as an eigenvalue problem for the frequencies $\omega_{n\ell m}$. Indeed, treating the QNM problem in terms of a PDE is crucial when the TEs do not decouple, for instance in the Kerr-Newman background~\cite{Dias:2015wqa}. In ref.~\cite{Dias:2015wqa}, however, the equation is not solved directly as an eigenvalue problem. By using a Newton-Raphson scheme, the QNMs are still found via a root-search algorithm requiring an initial guess for $\omega_{n\ell m}$. Besides, frequency-domain $m$-mode approaches have recently gained increasing attention in black hole perturbation theory. In the context of self-force calculations for instance~\cite{PanossoMacedo:2024pox}, $m$-mode hyperboloidal elliptic solvers have been shown to overcome limitations arising from the slow convergence of the spheroidal-harmonic $(\ell,m)$-mode sum due to the presence of a point particle. Similarly, in the study of QNMs, a hyperboloidal $m$-mode approach has recently been employed for calculating the pseudospectra of the Klein-Gordon equation for a massless field on the Kerr spacetime~\cite{Cai:2025irl}.

\smallskip
These recent advances underscore the need for a comprehensive study of this methodology for QNM calculations, which we present in this work. For this purpose, the next section reviews the Kerr spacetime, focusing on its underlying hyperboloidal formulation. Section \ref{sec:Kerr} also introduces the Teukolsky equation and its hyperboloidal $m$-mode scheme. The numerical methods are then summarised in Sec.\ref{sec:Numerics}, while Sec.\ref{sec:Results} presents our results, followed by concluding remarks in Sec.~\ref{sec:Conclusion}. We adopt geometrical units with $G = c = 1$.

\section{The Kerr spacetime}\label{sec:Kerr}
We begin with a brief review of the Kerr spacetime, focusing on the construction of the hyperboloidal foliation in the minimal gauge\cite{PanossoKERR}, and the resulting form of the Teukolsky equation.  
\subsection{Coordinates: from Boyer-Lindqst to hyperboloidal}\label{sec:BL_to_Hyp}

In terms of the traditional Boyer-Lindqst coordinates $x^\mu = (\BLCoord)$ the line element of the Kerr metric is written as
\bea
\label{KerrMetricBL}
ds^2 &=& -f dt^2  -  \dfrac{4Mar}{\Sigma} \sin^2\theta dtd\phi  + \dfrac{\Sigma}{\Delta} dr^2 \nn \\
&& +  \Sigma d\theta^2 + \sin^2\theta\left(\Sigma_0 + \dfrac{2Ma^2r}{\Sigma} \sin^2\theta\right)d\phi^2 ,
\eea
with
\bea
&&\Delta =  r^2 -2Mr + a^2 = (r-\rh )(r-\rc), \quad f = 1 - \dfrac{2Mr}{\Sigma},  \label{eq:Delta}  \\
\label{eq:Sigmas}
&&\Sigma = r^2 + a^2\cos^2\theta, \quad  \Sigma_0 = \left.\Sigma\right|_{\theta=0} = r^2 + a^2.
\eea
The black hole's mass and angular momentum are parametrized by $M$ and $a$, respectively. The zeros of the function $\Delta$ in Eq.~\eqref{eq:Delta} define the the event ($\rh$) and Cauchy ($\rc$) horizons
\bea
\dfrac{\rh}{M} =  1 + \sqrt{1 - \dfrac{a^2}{M^2}}, \quad \dfrac{\rc}{M} =  1 - \sqrt{1 - \dfrac{a^2}{M^2}}.
\eea

We employ an alternative parametrization for the spacetime, which uses the event horizon radius $\rh$ as length scale (as opposed to the mass $M$). In this way, the dimensionless black hole spin parameter $\kappa$ is equivalently defined by
\beq
\kappa = \dfrac{a}{\rh}=\dfrac{a}{M + \sqrt{M^2 - a^2}} \; \Longleftrightarrow \; \kappa^2 = \dfrac{\rc}{\rh}.
\eeq
As a consequence, the mass parameter reads 
\beq
M = \dfrac{\rh}{2} \left(1+\kappa^2\right).
\eeq

A hyperboloidal folitation $\bar x^\mu = (\HypCoord)$ then follows via the scri-fixing techinique\cite{Zenginoglu:2007jw}. We work in the so-called minimal gauge\cite{PanossoKERR} with the notation from \cite{PanossoSphSym,Minucci:2024qrn}, for which the transformation from BL into hyperboloidal coordinates reads
\bea
\label{eq:HypTrasfo}
t = \rh \bigg( \tau - H(\sigma) \bigg), \quad
r = r_o + \dfrac{\rh - r_0}{\sigma},  \quad
\cos \theta = x, \quad
\phi = \varphi - \bar \chi(\sigma).
\eea

The radial compactification in Eq.~\eqref{eq:HypTrasfo} keeps the minimal structure needed to map the infinitely far wave zone to finite coordinate values. Thus, $\sigma = 0$ corresponds to future null infinity, while the BH horizon $r = \rh$ is conveniently set to $\sigma_{\rm h} = 1$. Such a compactification allows for an arbitrary radial offset $r_o$, which influences the resulting coordinate location for the Cauchy horizon as $r =\rc $ gets mapped into 
\beq
\sigma_{\rm c} = \dfrac{\rh - r_o}{\rc - r_o} 
		       = \dfrac{1- \rho_o}{\kappa^2 - \rho_o},
\eeq
with\footnote{The notation differs slightly from Refs.~\cite{PanossoKERR,PanossoSphSym}. The parameter $\rho_o$ here, corresponds to $\rho_1$ in \cite{PanossoKERR,PanossoSphSym}.} $\rho_o = \dfrac{r_o}{\rh}$. As detailed in refs. \cite{PanossoKERR}, we consider two options:
\begin{itemize}
    \item {\bf The radial fixing gauge:}
    
    The straightfoward choice $r_o=0$ ($\rho_o=0$) implies $\sigma_{\rm c} = \kappa^{-2}$. In this way, $\sigma_{\rm c}$ depends on the rotation BH rotation parameter $\kappa$, going smoothly from the Schwarzschild singularity $\sigma_{\rm c}\rightarrow \infty$ as $\kappa \rightarrow 0$ to the the extremal value $\sigma_{\rm c}=\sigma_+-1$ as $\kappa \rightarrow 1$. This option corresponds to the one employed in ref.~\cite{Ripley:2022ypi}.

    \item {\bf The Cauchy horizon fixing gauge:}
    
    The choice $r_o=\rc$ ($\rho_o=\kappa^2$) fixes $\sigma_{\rm c}\rightarrow \infty$ for all values of $\kappa$. In this way, the hypersurfaces $\tau = $ constant do not cross the Cauchy horizon\cite{Minucci:2024qrn}, and the extremal limit provides a discontinuous transition into the Kerr near-horizon geometry\cite{PanossoKERR}. This option corresponds to the spacetime counterpart of Leaver approach~\cite{Leaver}.
\end{itemize}

\begin{figure}[b!]
    \centering
    \includegraphics[width=0.5\columnwidth]{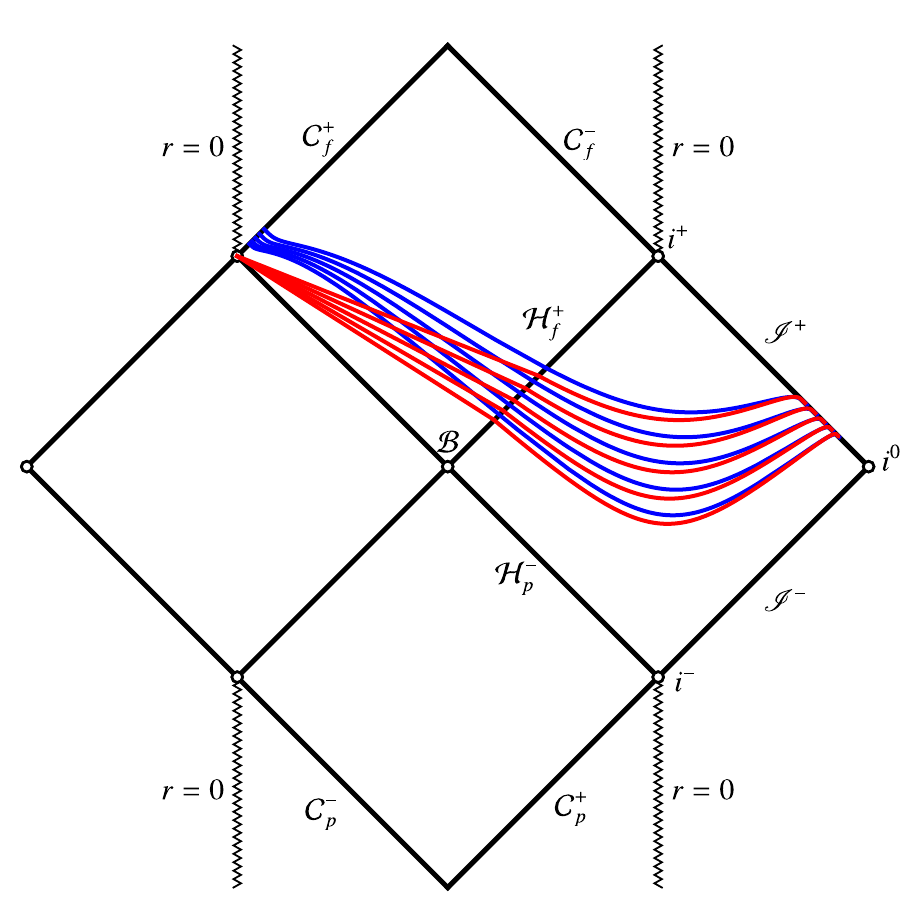}
    \caption{Carter-Penrose diagram for the Kerr spacetime, illustrating hyperboloidal slices in the radial-fixing minimal gauge (blue) and the Cauchy-horizon-fixing minimal gauge (red). In the former, the coordinate position of the Cauchy horizon is parametrised by the black hole spin parameter via $\sigma_{\rm c}=\kappa^{-2}$, ensuring that the $\tau=$ constant hypersurfaces foliates $C_{\rm f}^+$ continuously, with a smooth transition in the extremal Kerr limit $\kappa=1$. In contrast, the latter gauge fixes $\sigma_{\rm c} \rightarrow \infty$, making it independent of $\kappa$. Here, the $\tau=$ constant hypersurfaces accumulate at the singular point connecting $C_{\rm f}^+$ to $r=0$ and $i^+$, and the limit $\kappa \to 1$ has a discontinuous transition into the near-horizon geometry. Adapted from \cite{Minucci:2024qrn}. }
    \label{fig:Penrose}
\end{figure}

The height function $H(\sigma)$ in the minimal gauge, and the azimuthal function $\bar \chi(\sigma)$ reads~\cite{PanossoKERR,PanossoSphSym,Minucci:2024qrn}
  \beq
  \label{eq:H_MinGaug}
  H(\sigma) = -\dfrac{1 - r_o/\rh}{\sigma} + \dfrac{2M}{\rh} \ln \sigma + \dfrac{1}{2 \rh \varkappa_h } \ln(1- \sigma) + \dfrac{1}{2 \rh \varkappa_c } \ln\left(1- \dfrac{\sigma}{\sigma_c}\right),
  \eeq
\beq
\bar \chi(\sigma) = \dfrac{\Omega_h}{2 \varkappa_h } \ln(1- \sigma) + \dfrac{\Omega_c}{2  \varkappa_c } \ln\left(1- \dfrac{\sigma}{\sigma_c}\right).
\eeq
with $\varkappa_{h,c} = (4M)^{-1} \left( 1 - a^2/r_{h,c}^2 \right)$ and $\Omega_{h,c} = a \left( 2M r_{h,c}\right)^{-1}$, respectively, the horizons' surface gravity, and angular velocities. In terms of the dimensionless parameter $\kappa$ they read
\bea
\varkappa_{\rm h} =  \dfrac{1-\kappa^2}{2\rh (1+\kappa^2)}, &\quad& \varkappa_{\rm c} =  - \dfrac{1-\kappa^2}{2\rh \kappa^2 (1+\kappa^2)}, \\
\Omega_{\rm h} = \dfrac{\kappa}{\rh (1+\kappa^2)}, & \quad & \Omega_{\rm c} = \dfrac{1}{\rh \kappa (1+\kappa^2)}.
\eea
Finally, the mapping in the polar angle $\theta(x)$ in Eq.~\eqref{eq:HypTrasfo} is a convenient transformation for the upcoming numerical treatment of the Teukolsky equation.

\smallskip
The explicit form of the resulting line element \eqref{KerrMetricBL} is not of primary relevance here. However, it is worth noting that the hyperboloidal transformation \eqref{eq:HypTrasfo} naturally induces a conformal rescaling of the metric \eqref{KerrMetricBL}, given by
\beq
\label{eq:conformal_metric}
g_{ab} = \Omega^{-2} \bar g_{ab}, \quad \Omega = \dfrac{\sigma}{\rh},
\eeq
where $\bar g_{ab}$ remains finite and regular. This transformation highlights the advantage of working with conformally compactified spacetimes.

\smallskip
Accordingly, it is more insightful to represent the structure of Kerr spacetime using a Carter-Penrose diagram (fig.~\ref{fig:Penrose}), which illustrates the hyperboloidal slices associated with the radial-fixing (blue) and Cauchy horizon-fixing (red) gauges \cite{Minucci:2024qrn}. While both foliations provide valid hyperboloidal slicings in the exterior black hole region, their behavior at the Cauchy horizon differs significantly. In the radial-fixing gauge, the constant-$\tau$ surfaces smoothly cross $C_{\rm f}^+$, making $\sigma_{\rm c}$ a regular surface. Conversely, in the Cauchy horizon-fixing gauge, these slices accumulate at the singular point connecting $C_{\rm f}^+$ to $i^+$ and the central singularity, leading to a qualitatively distinct asymptotic structure.

\smallskip
The next section reviews the effects of the coordinate transformation \eqref{eq:HypTrasfo} in the Teukolsky equation.

\subsubsection{Teukolsky equation} \hfill \\
Traditionally, the Teukolsky equation is expressed in BL coordinates $x^\mu = (\BLCoord)$. In the absence of source terms (vacuum case), the Teukolsky master function $\spinPsi{\spin}(\BLCoord)$ satisfies
\bea
\label{TeukolskyEqBL}
&&0 = \left(\dfrac{(\Sigma_0)^2}{\Delta} - a^2 \sin^2\theta \right)\partial_{tt}^2\spinPsi{\spin} \ + \ \dfrac{4Mar}{\Delta}\partial_{t\phi}^2\spinPsi{\spin} \ + \ \left(\dfrac{a^2}{\Delta} - \dfrac{1}{\sin^2\theta} \right)\partial_{\phi \phi}^2\spinPsi{\spin}  \nn \\
&& - \ \Delta^{-s} \partial_r (\Delta^{s+1} \partial_r \spinPsi{\spin}) \ - \ 2s\left(\dfrac{M(r^2 - a^2)}{\Delta} - (r + ia\cos\theta) \right) \partial_t \spinPsi{\spin} \\
&&- 2s\left(\dfrac{a(r - M)}{\Delta} + i\dfrac{\cos\theta}{\sin^2\theta} \right) \partial_\phi \spinPsi{\spin} \ - \ \dfrac{1}{\sin\theta} \partial_\theta (\sin\theta \partial_\theta \spinPsi{\spin}) \ + \ s(s \cot^2\theta - 1)\spinPsi{\spin}. \nn 
\eea
The parameter $\spin$ characterizes the spin weight of the perturbing field, with $\spin=0$ corresponding to scalar field perturbations, $\spin = \pm 1$ describing electromagnetic waves, and $\spin = \pm 2$ representing gravitational waves. Typically, the Teulkolsky equation \eqref{TeukolskyEqBL} is re-formulated in the frequency domain via a Fourier (or Laplace) transformation, leading to a separation of variables into ordinary differential equations (ODEs) governing the radial and angular dependencies. The quasi-normal mode (QNM) problem is then formulated by imposing outgoing boundary conditions on the radial ODE.

\smallskip
We depart from the traditional approach by first introducing hyperboloidal coordinates \eqref{eq:HypTrasfo}, which inherently enforce the correct boundary conditions geometrically. To further ensure the regularity of the transformed function for $(\sigma, x) \in [0,1]\times[-1,1]$, we redefine the original Teukolsky master function as~\cite{ZenginogluGeoBH,PanossoKERR}

\beq
\label{eq:TeukFunc_BL_HypRadial}
\spinPsi{\spin}(x^a) = \Omega \, \Delta^{-\spin}  \, \left(1 + x\right)^{\delta_1/2}\left(1 - x\right)^{\delta_2/2} \,\, \spinHyperPsi{\spin}(\bar x^a).
\eeq
Apart from the direct coordinate transformation $x^a(\bar x^a)$, the factor $\Delta^{-\spin}$ arises from boosting the Kinnersley null tetrad—originally used to define $\spinPsi{\spin}$—ensuring that the resulting hyperboloidal null tetrad basis remains regular at the black hole horizon \cite{Minucci:2024qrn}. The additional factor $\Omega$ ensures compatibility with the formulation of the Teulkolsky equation directly from the the conformal spacetime\eqref{eq:conformal_metric}~\cite{Gasperin2025}. Finally, the terms $(1\pm x)^{\delta_{1,2}/2}$, where $\delta_1 = \vert m - \spin \vert$ and $\delta_2 = \vert m + \spin \vert$, imposes the function’s regularity at the symmetry axis $x = \pm1$ (or equivalently $\sin\theta = 0$)~\cite{PanossoKERR}.
 
 \smallskip
The desired $m$-mode hyperboloidal Teukolsky equation follows from a transformation into the frequency domain, applied to the time and azimuthal angle:
\beq
\label{eq:Fourier_hyperRadial}
\spinHyperPsi{\spin}(\HypCoord) = \dfrac{1}{2 \pi} \int_{-\infty }^{\infty } \sum_{m=-\infty}^{\infty} e^{-i \bar \omega \tau} e^{i m \bar \phi} \spinHyperPhi{\spin} \, {\rm d} \bar\omega.
\eeq
In this expression, $\bar \omega$ is a dimensionless frequency parameter, related to the physical frequency $\omega$ from the standard ansatz $\sim e^{- i \omega t}$ via the rescaling $\bar \omega = \rh \omega$. This follows naturally from the coordinate transformation \eqref{eq:HypTrasfo}, where $\rh$ sets the characteristic length scale. Hence, the resulting equation assumes the form (with $s = - i \bar \omega$)
\beq
\label{eq:HypTeuk_PDE}
 {}_{\spin}\bar{\cal D}_{m;\bar \omega} \left[ \spinHyperPhio{\spin}  \right] = 0, \quad 
  {}_{\spin}\bar{\cal D}_{m;\bar \omega} =  {\rm L}_1 + s {\rm L}_2 -w(\sigma, x) s^2,
\eeq
with ${\rm L}_1$ and ${\rm L}_2$, respectively, second and first order differential operators acting only on the spatial coordinates $(\sigma, x)$ represented by
\bea
\label{eq:L1}
{\rm L}_1 = \alpha_2(\sigma)\partial^2_{\sigma\sigma} + \alpha_1(\sigma)\partial_{\sigma} + \alpha_0(\sigma) + \Gamma_2(x) \partial^2_{xx}  + \Gamma_1(x) \partial_{x}, \\
\label{eq:L2}
{\rm L}_2 = \beta_1(\sigma)\partial_{\sigma} +  \beta_0(\sigma, x).
\eea
Then, the QNM eigenvalue problem results from the introduction of an auxiliary function $\spinHyperUpsilono{\spin} = s \, \spinHyperPhio{\spin}$,  which yields
\beq
\label{eigenpb}
{\rm L} u = s u, \quad {\rm L} = \left(
\begin{array}{cc}
0	& 1 \\
w^{-1} {\rm L}_1	&w^{-1} {\rm L}_2
\end{array}
\right), \quad
u = \left(
\begin{array}{c}
\spinHyperPhio{\spin} \\
\spinHyperUpsilono{\spin}
\end{array}
\right)
\eeq

For the radial fixing gauge ($\sigma_c = \kappa^2$), the functions appearing above read
\bea
w(\sigma, x) &= 4 (\sigma +1)+\kappa ^2 \bigg(-4 \left(\kappa ^2+1\right)^2 \sigma ^2 +4 \kappa ^2    \sigma  +x^2+7 \nn \\
 & +4 \left(\kappa ^4+2 \kappa ^2+2\right) \bigg),\\
\alpha_2(\sigma) &= \sigma ^2 (1-\sigma )  \left(1-\kappa ^2 \sigma \right) ,\\
\alpha_1(\sigma) &= \sigma  \left(4 \kappa ^2 \sigma ^2-\left(\kappa ^2+1\right) \sigma  (\spin+3)-2 i \kappa  m \sigma +2 (\spin+1)\right),\\
\alpha_0(\sigma) &= \frac{1}{2} \bigg(-\delta_1 (\delta_2+1)-\delta_2+4 \kappa ^2 \sigma ^2-2 \left(\kappa ^2+1\right) \sigma 
   (\spin+1) \nn \\
   & -m^2-4 i \kappa  m \sigma +\spin (\spin+2)\bigg),\\
\beta_1(\sigma) &= 2 \sigma ^2 \left(2 \kappa ^4 (\sigma -1)+\kappa ^2 (2 \sigma -3)-2\right)+2 ,\\
\beta_0(\sigma,x) &= -2 \sigma  \left(\kappa ^2 \left(\kappa ^2 (\spin+2)+2 \spin+3\right)+\spin+2\right)+6 \left(\kappa ^4+\kappa ^2\right) \sigma ^2  \nn \\
&-2 i m
   \left(2 \left(\kappa ^2+1\right) \kappa  \sigma +\kappa \right)+2 \spin \left(\kappa ^2-i \kappa  x+1\right) .
\eea
In the Cauchy horizon fixing gauge $(\sigma_{\rm c}\rightarrow \infty)$, they read
\bea
w(\sigma, x) &= \frac{4 \left(\kappa ^2+1\right)^2 \sigma }{1-\kappa ^2}+\kappa ^2 \left(x^2+3\right)+4,\\
\alpha_2(\sigma) &= \sigma ^2 (1-\sigma ),\\
\alpha_1(\sigma) &= \frac{\sigma  \left(\kappa ^2 (2 (\spin+1)-\sigma  (\spin+3))+2 i \kappa  m \sigma +\sigma  (\spin+3)-2 (\spin+1)\right)}{\kappa
   ^2-1},\\
\alpha_0(\sigma) &= -\frac{1}{2} \bigg( \delta_1 (\delta_2+1) + \delta_2 + m^2 - \frac{4 i \kappa  m \sigma }{\kappa ^2-1} + 2 \sigma (\spin+1) \nn \\
&- \spin (\spin+2)\bigg),\\
\beta_1(\sigma) &= \frac{-2 \kappa ^4 (\sigma -1)^2+2 \kappa ^2 ((\sigma -2) \sigma +2)+4 \sigma ^2-2}{\kappa ^2-1},\\
\beta_0(\sigma, x) &= \frac{2}{\kappa ^2-1} \bigg(-\left(\kappa ^4 (\sigma -1) (\spin+1)\right)+\kappa ^2 (\sigma -1) +(\sigma -1) \spin +2 \sigma \nn \\
&+i m \left(\kappa ^3 (2 \sigma -1) + 2 \kappa  \sigma +\kappa  \right)  -i \kappa ^3 x \spin+i \kappa  x \spin\bigg).
\eea
Moreover, we have in both options
\bea
\Gamma_2(x) &= 1-x^2,\\
\Gamma_1(x) &=  \delta_1-\delta_2-x(\delta_1+\delta_2+2).
\eea
The first thing to note is that all these functions remain regular within the domain $(\sigma, x) \in [0,1] \times [-1,1]$. However, their behavior differs in the $\kappa \to 1$ limit, depending on whether we use the radial fixing gauge or the Cauchy horizon fixing gauge. In the former, all functions are well-defined, whereas in the latter, they exhibit a divergence of the form $\sim (1-\kappa^2)^{-1}$. This distinction arises from the fundamental properties of these foliations: while the radial fixing gauge smoothly extends to the extremal Kerr geometry, the Cauchy horizon gauge undergoes a discontinuous transition into the near-horizon region.

\smallskip

One could, in principle, proceed by separating variables as $\spinHyperPhio{\spin}(\sigma, x) = \spinHypRo{\spin} \spinHypSo{\spin}$, yielding ODEs for $\spinHypRo{\spin}$ and $\spinHypSo{\spin}$, coupled by the separation constant $\spinlambda{\spin}{\bar \omega}$ and the frequency $\bar \omega$. As expected, $\spinHypSo{\spin}$ corresponds to spin-weighted spherical harmonics. The radial function $\spinHypRo{\spin}$ remains regular at the QNMs, allowing for an analysis similar to Leaver’s method \cite{Leaver}, where an ansatz of the form $\spinHypRo{\spin} = \sum_{k=0}^{\infty} a_k (1-\sigma)^k$ leads to a recurrence relation for the sequence $\{a_k\}_{k=0}^{\infty}$. The QNMs then correspond to the values of $\omega_n$ for which this sequence forms a minimal solution.

\smallskip

In fact, the Cauchy horizon fixing gauge reproduces exactly Leaver’s continued fraction method. A similar approach could be applied to the radial fixing gauge, but the recurrence relation for $\{a_k\}_{k=0}^{\infty}$ is no longer restricted to three terms, and therefore, the continued fraction method must be adapted accordingly.  Moreover, this method is only valid for $\kappa^2 < 1/2$, where the Cauchy horizon is located at $\sigma_{\rm c} > 2$, placing it outside the convergence radius of the Taylor series \cite{PanossoKERR,PanossoMacedo:2018hab}. An alternative strategy is to approximate $\spinHypRo{\spin}$ via spectral decomposition using Chebyshev polynomials, as done in \cite{Ripley:2022ypi} and refined in \cite{Zhu:2023mzv}. These works had indeed the ODE for $\spinHypRo{\spin}$ in the radial fixing gauge. We recall that the radial and angular equation must be solved together as they are coupled by $\spinlambda{\spin}{\bar \omega}$ and $\bar \omega$. For $\spinHypSo{\spin}$, Leaver also lays out a strategy based on a Taylor expansion around one boundary, and solved using the continued fraction methods. However, ref.~\cite{Cook:2014cta} reports that a Chebyshev spectral method is more efficient for solving the angular equation as well.

\smallskip
Thus, in this work, we also adopt a spectral decomposition approach based on Chebyshev polynomials, but instead of applying it to the individual radial and angular functions alone, we directly approximate the full 2D function $\spinHyperPhio{\spin}(\sigma, x)$.

\section{Numerical Methods}\label{sec:Numerics}
We discretize the eigenvalue problem Eq.~\eqref{eigenpb} by approximating the hyperboloidal $m$-mode master function as
\beq
\label{eq:cheby_apprxo}
\spinHyperPhio{\spin}(\sigma(\chi), x) \approx \sum_{k_1=0}^{N_1} \sum_{k_2=0}^{N_2} c_{k_1,k_2} T_{k_1}(\chi) T_{k_2}(x),
\eeq
with $T_k(\xi) = \cos\left[ k \arccos(\xi) \right]$ being the Chebyshev polynomials of the first kind, naturally defined in the domain $\xi \in[-1,1]$. The Chebyshev coefficients $c_{k_1,k_2}$ are fixed via a collocation method, which imposes that the approximation is exact at a set of grid points. We work with the Lobatto-Chebyshev nodes
\bea
\label{eq:nodes_1}
&\chi_{i_1} &= \cos \left(  \frac{i_1 }{N_1}\pi \right), \quad  i_1 = 0, \dots, N_1, \quad n_1 = N_1 +1; \\
\label{eq:nodes_2}
&x_{i_2} &= \cos \left(  \frac{i_2 }{N_2}\pi \right), \quad  i_2 = 0, \dots, N_2, \quad n_2 = N_2 +1,
\eea
with $n_1$ and $n_2$ the total number of grid points along each individual direction.

\smallskip
Justifying the transformation in Eq.~\eqref{eq:HypTrasfo}, the angular coordinate $x = \cos\theta \in[-1,1]$ is directly suited to the spectral representation \eqref{eq:cheby_apprxo}. The radial coordinate $\sigma \in [0,1]$, however, must be mapped to $\chi \in [-1,1]$ via
\bea
\label{eq:AnMR}
    \sigma(\chi) = \dfrac{1 + \chi }{2}.
 \eea

\smallskip
By differentiating Eq.~\eqref{eq:cheby_apprxo} with respect to $\chi$ or $x$ and evaluating it at the corresponding collocation points, derivatives $\partial_\xi$ along either direction, $\xi = \chi$ or $\xi = x$, are represented by the discrete differentiation matrix $\hat D_{\xi}$, with components
\begin{equation}
\label{eq:D_Lobatto}
\left(\hat D_{\xi}\right)_{ij} = 
\left\{
\begin{array}{cc}
      \dfrac{k_i (-1)^{i-j}}{k_j (x_i-x_j)} & i \neq j, \\
      \\
      -\dfrac{x_j}{2 (1-x_j^2)} & 0 < i=j < N, \\
      \\
      \dfrac{2N^2+1}{6} & i=j=0, \\
      \\
      -\dfrac{2N^2+1}{6} & i=j=N,
\end{array} \right., 
\quad
    k_i = 
\left\{
\begin{array}{cc}
        2 & i=0,N, \\
        1 & i \neq 0,N.        
    \end{array} \right.
\end{equation}
Then, second derivatives follow directly by a matrix-matrix multiplication $\hat D_{\xi\xi} = \hat D_{\xi} \cdot \hat D_{\xi}$. 

\smallskip
While these matrices are directly applicable along the angular direction $x$, i.e. $\hat D_x = \hat D_\xi$, the corresponding matrix $\hat D_\sigma$ necessary to perform numerical  derivatives along the radial direction must be adapted according to the re-scalling \eqref{eq:AnMR}. With $\hat D_\chi = \hat D_\xi$, the differentiation matrix along the radial direction re-scales as 
\bea
\label{eq:Dsigma}
\hat D_\sigma = 2  \hat D_{\chi}, \quad \hat D_{\sigma\sigma} = \hat D_\sigma \cdot \hat D_\sigma,
\eea

The discretisation scheme described above is restricted to a representation along the individual coordinates $\sigma$ and $x$. Thus, the matrices $\hat D_\sigma$ and $\hat D_x$ have sizes $n_1\times n_1$ and $n_2\times n_2$, respectively. The $2D$ problem arises from a tensorial product of the two spatial dimensions and therefore the discrete eingenfunction $\vecspinHyperPhi{\spin}$ results from flattening the tensorial product into a vector of size $n_{\rm total} = n_1 \times n_2$
\bea
&\vecspinHyperPhi{\spin} = \sum_{k_1=0}^{N_1}\sum_{k_2=0}^{N_2} c_{k_1, k_2} \flatten{  \vec{T}_{k_1}\otimes \vec{T}_{k_2} }, \\
&\left(\vec{T}_{k_1}\right)_i = T_{k_1}(\chi_i), \quad \left(\vec{T}_{k_2}\right)_i = T_{k_2}(x_i),
\eea
In this way, one needs the tensorial product extension of all its elements of eq.~\eqref{eigenpb} to construct its discretised version. For that purpose, we introduce the grid vectors
\beq
\label{eq:tensor_grids}
\vec{\boldsymbol{\sigma}} = \flatten{ \vec{\sigma}\otimes \vec{1}_{n_2} }, \quad \vec{\boldsymbol{x}} = \flatten{\vec{1}_{n_1} \otimes\vec{x}},
\eeq
with $\vec{\sigma}$ and $\vec{x}$ the grid points associated with the nodes \eqref{eq:nodes_1} and \eqref{eq:nodes_2}, respectively. Besides, $\vec{1}_{n_1}$ and $\vec{1}_{n_2}$ are constant unit vector of size $n_1$ and $n_2$, respectively. With the help of eq.~\eqref{eq:tensor_grids}, we obtain the discrete version of the functions defining the operators ${\rm L}_1$ and ${\rm L}_2$ in eqs.~\eqref{eq:L1} and \eqref{eq:L2} as
\bea
\vec{\boldsymbol w} = w(\vec{\boldsymbol{\sigma}},\vec{\boldsymbol{x}}), \quad \vec{\boldsymbol \alpha}_2 = \alpha_2(\vec{\boldsymbol{\sigma}}),  \quad \vec{\boldsymbol \alpha}_1 = \alpha_1(\vec{\boldsymbol{\sigma}}),  \quad \vec{\boldsymbol \alpha}_0 = \alpha_0(\vec{\boldsymbol{\sigma}}),  \\
\vec{\boldsymbol \beta}_1 = \beta_1(\vec{\boldsymbol{\sigma}}), \quad \vec{\boldsymbol \beta}_0 = \beta_0(\vec{\boldsymbol{\sigma}},\vec{\boldsymbol{x}}), \quad \vec{\boldsymbol \Gamma}_2 = \Gamma_2(\vec{\boldsymbol{x}}), \quad \vec{\boldsymbol \Gamma}_1 = \Gamma_1(\vec{\boldsymbol{x}}).
\eea
Similarly, the $1D$ differentiation matrices along $\sigma$ and $x$ are expanded into $2D$ matrices with size $n_{\rm total}\times n_{\rm total}$  via the tensor product
\bea
\hat {\boldsymbol D}_\sigma = \hat D_{\sigma}\otimes{\mathbb 1}_{n_2}, \quad \hat {\boldsymbol D}_{\sigma\sigma} = \hat D_{\sigma\sigma}\otimes{\mathbb 1}_{n_2} \\
\hat {\boldsymbol D}_x = {\mathbb 1}_{n_1}\otimes \hat D_{x} , \quad \hat {\boldsymbol D}_{xx} = {\mathbb 1}_{n_2}\otimes \hat D_{xx},
\eea
with ${\mathbb 1}_{n}$ the identity matrix with size $n \times n$. With the above elements, we obtain the discrete operator
\beq
\label{eq:disc_L}
\hat{\boldsymbol L} = \left(
\begin{array}{cc}
{\mathbb 0}_{n_{\rm total}} & {\mathbb 1}_{n_{\rm total}} \\
\vec{\boldsymbol w}^{-1}\circ \hat {\boldsymbol  L}_1 & \vec{\boldsymbol w}^{-1}\circ \hat {\boldsymbol  L}_2
\end{array}
\right),
\eeq
where ${\mathbb 0}_{n}$ is the zero matrix with size $n \times n$, and 
\bea
 \hat {\boldsymbol  L}_1 = \vec{\boldsymbol \alpha}_2 \circ \hat {\boldsymbol D}_{\sigma\sigma} +  \vec{\boldsymbol \alpha}_1 \circ \hat {\boldsymbol D}_{\sigma}  +  \vec{\boldsymbol \alpha}_0 \circ {\mathbb 1}_{n_{\rm total}}  + \vec{\boldsymbol \Gamma}_2 \circ \hat {\boldsymbol D}_{xx} + \vec{\boldsymbol \Gamma}_1 \circ \hat {\boldsymbol D}_{x}, \\
  \hat {\boldsymbol  L}_2 =   \vec{\boldsymbol \beta}_1 \circ \hat {\boldsymbol D}_{\sigma}  +  \vec{\boldsymbol \beta}_0 \circ {\mathbb 1}_{n_{\rm total}}.
\eea
In the eq.~\eqref{eq:disc_L}, we have abused the vector notation when expressing $\vec{\boldsymbol w}^{-1}$, which should be understood under definition through its components
$
\left( \vec{\boldsymbol w}^{-1}\right)_i = 1/ \boldsymbol w_i.
$
 Finally, the operator $\circ$ denotes the element-wise (Hadamard) product, defined ---for any given vector $\vec v$ and matrix $\hat M$ --- by
$
    \left(\vec{v}\circ \hat M\right)_{ij} = v_i \, M_{ij}.
$

With the discretised operator, one then extracts the QNM frequencies and regular functions from the eingevalues and eigenvectors of $\hat{\boldsymbol L}$.

\section{Results}\label{sec:Results}
With the numerical scheme adapted to solving the hyperboloidal TEs as a PDE, we now present the results obtained by treating the problem as a direct eigenvalue computation. Our primary focus is on the gravitational case with $\spin = -2$, although the framework is equally applicable to other cases. Furthermore, the hyperboloidal transformation defined in Eq.~\eqref{eq:HypTrasfo} introduces $\rh$ as the characteristic length scale of the problem, thereby setting the natural scaling for the eigenvalues $s = -i \rh \omega$ associated with $\hat{\boldsymbol L}$. To facilitate direct comparison with values available in the literature, we normalize our results by the black hole mass, i.e.,
\beq
M \omega = \dfrac{i \,s}{2(1+\kappa^2)}.
\eeq

The numerical discretization of the PDE introduces the parameters $n_1$ and $n_2$, cf. Eqs.~\eqref{eq:nodes_1} and \eqref{eq:nodes_2}, which control the number of grid points along the radial and angular directions, respectively. For a fixed value of $m$, each truncation parameter determines the total number of QNMs that can be obtained, associated with the polar index $\ell$ and the overtone index $n$. In the angular direction, the maximum number of $\ell$ modes that the solver can recover is directly related to the truncation parameter via $\ell_{\rm max} = n_2 - |\spin|$. In the radial direction, however, the treatment requires more care.

\smallskip
As reported in Ref.\cite{jaramillo2021pseudospectrum} for the $1D$ problem, QNMs constitute only a subset of the total eigenvalues associated with the discrete operator $\hat{\boldsymbol L}$. The QNM values converge exponentially to a fixed limit as the truncation parameter $n_1$ increases. However, a significant number of non-converging eigenvalues remain. Among these spurious eigenvalues, one finds values along $\mathrm{Re}(s) < 0$, which are commonly interpreted as the discretized representation of the well-known branch cut present in the operator’s Green’s function. Additionally, the operator exhibits spurious eigenvalues forming an arc at high frequencies, which are typically discarded.\footnote{Interestingly, Ref.\cite{Besson:2024adi} reports that a complete description of the time evolution can be achieved when the non-converging eigenvalues along the branch cut are retained in their formalism.} 

\smallskip
Beyond the non-converging values discussed above, the $2D$ representation of the QNM problem introduces additional spurious eigenvalues, also observed in $1D$ formulation of rotating BH analogues~\cite{dePaula:2025fqt}.  These values cluster near the branch cut at $\mathrm{Re}(s) < 0$ but possess a small, nonzero imaginary component. Notably, they become increasingly dense as the black hole spin parameter $a/M$ approaches extremality. To systematically filter out these spurious eigenvalues within a given tolerance $\mathrm{TOL} \ll 1$, we solve the QNM eigenvalue problem for two different values of the radial truncation parameter, $n_1^{\rm low}$ and $n_1^{\rm high}$, while keeping $n_2$ fixed. We then apply a filtering criterion, retaining only those eigenvalues that satisfy $\left| 1 - \omega^{\rm low}/\omega^{\rm high} \right| < {\rm TOL}$. With a ${\rm TOL} \sim 10^{-3}$, one roughly retains a total of $n_{\rm overtones} \sim \sqrt{n_1}$ overtones. The eigenvalue problem is solved with \texttt{Wolfram Mathematica}, were we set the internal precision for float calculations as $10\times$\texttt{MachinePrecision}.

\subsection{QNMs and symmetry of the problem: new notation}
\begin{figure}[t!]
    \centering
    \includegraphics[width=0.8\columnwidth]{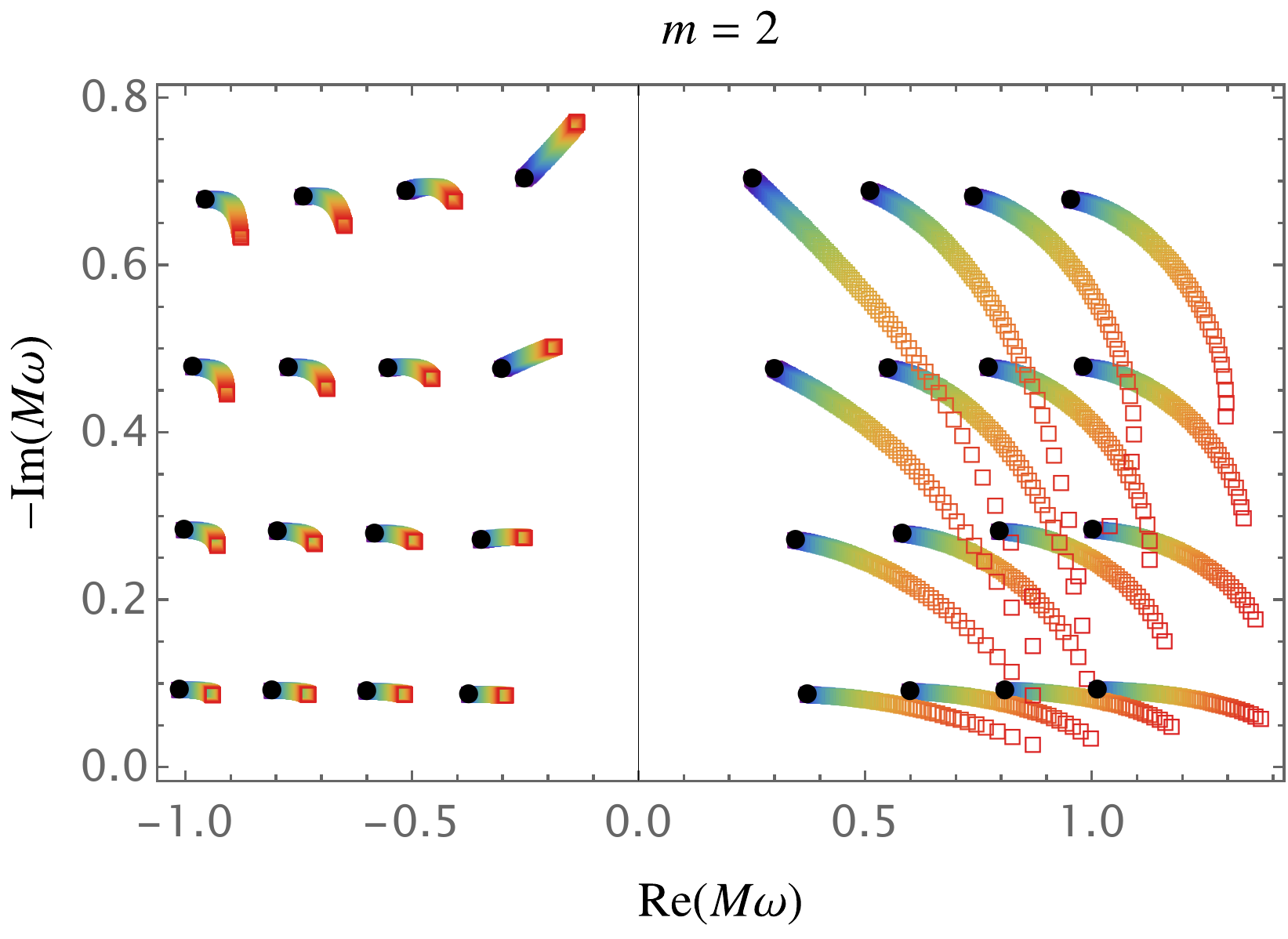}
    \caption{ QNM spectrum for spin $\spin = -2$ in the sector $m=2$, obtained using the hyperboloidal $m$-mode eigenvalue solver. Results for black hole spins ranging from $a/M = 0$ (black dots) to $a/M = 0.99$ (red squares). This method enables the simultaneous extraction of multiple QNMs—including higher harmonics and overtones—without the need for fine-tuned initial seeds in root-finding algorithms or prior assumptions about angular structure or mode classification. Modes emerge directly from the spectral problem, providing a robust and automated route to constructing the full QNM spectrum.}
    \label{fig:QNM_m2}
\end{figure}

Figure~\ref{fig:QNM_m2} displays the QNMs for $m=2$, obtained from the PDE eigenvalue problem for spin parameters $a/M \in [0,1]$. The converged QNM values were extracted after applying a filtering process with $n_2^{\rm high} = 11$, $n_1^{\rm high} = 55$, $n_2^{\rm low} = 10$ $n_1^{\rm low} = 50$, and a tolerance of ${\rm TOL} = 10^{-3}$. The figure highlights key properties of the eigenvalue solver. 

\smallskip
The black dots in Fig.~\ref{fig:QNM_m2} represent the Schwarzschild QNM values at $a/M = 0$, serving as an essential benchmark. In the spherically symmetric case, where the radial and angular equations fully decouple, the QNM eigenvalue problem has traditionally been solved for individual $(\ell, m)$ modes, with the eigenvalue solver yielding values for the overtone index $n = 0, 1, \dots$. Reformulating the Schwarzschild QNM eigenvalue problem as an $m$-mode PDE reveals not only a sequence of eigenvalues increasing in $\left| {\rm Im} (M \omega) \right|$ but also additional values spreading toward higher $\left| {\rm Re} (M \omega) \right|$. Since there is no explicit angular equation defining the index $\ell$, an alternative labeling is required for the eigenvalues appearing along $\left| {\rm Re} (M \omega) \right|$. Naturally, defining $\ell$ such that
\beq
\left| {\rm Re} (\omega_{n \, |\spin| m}) \right| < \left| {\rm Re} (\omega_{n \, |\spin|+1 \, m}) \right| < \left| {\rm Re} (\omega_{n \, |\spin|+2 \, m}) \right| \cdots
\eeq
confirms that the truncation parameter $n_2$ directly determines the different well-known $\ell$ modes, with the total number of available modes following the relation $\ell = n_2 - |\spin|$.

\smallskip
For $a/M > 0$, the degeneracy between the eigenvalues $\omega_{n\ell m}$ and $-\omega^\ast_{n\ell m}$ is broken. In the Schwarzschild case, both $\omega_{n\ell m}$ and $-\omega^\ast_{n\ell m}$ belong to the eigenvalue spectrum for a given $m$ (with the eigenvalues actually being independent of $m$). However, as the black hole rotation increases, the values $-\omega^\ast_{n\ell m}$ no longer belong to the spectrum for this particular $m$. Additionally, a clear distinction arises between prograde and retrograde modes, satisfying ${\rm sign} \left( {\rm Re} \, \omega \right) = {\rm sign} (m)$ and ${\rm sign} \left( {\rm Re} \, \omega \right) = -{\rm sign} (m)$, respecvelty. The black hole’s rotation parameter has a stronger impact on prograde modes, leading to a more significant frequency shift. Moreover, the symmetries of the equation reproduce the expected relation
\beq
\label{eq:MirrorMode}
\omega_{n \ell m} = -\omega_{n \ell \,-m}^\ast.
\eeq

\smallskip
The properties described above are well-documented in the literature. However, our approach offers a key advantage: it enables the computation of the complete QNM spectrum for a fixed $m$ at once. This eliminates the need for a root-finding algorithm to obtain each individual $\omega_{n \ell m}$. Since root-finding algorithms only converge when the initial seed is sufficiently close to the true root, they require a well-informed choice of initial guesses. In contrast, the eigenvalue solver provides QNM values directly, without any prior assumptions. In fact, using the eigenvalue solver with relatively low resolution is a practical strategy for generating initial seeds for more efficient root-finding techniques, as implemented in the \texttt{Black Hole Toolkit} for QNMs in Schwarzschild spacetime~\cite{BHPToolkit}.

\smallskip
Beyond computational advantages, visualizing the full QNM spectrum for a given $m$ sector also helps clarify a common terminology in the literature. Because root-finding methods require an initial guess, it has become standard practice to restrict QNM searches to the half-plane ${\rm Re}(\omega) > 0$. This means that, for a given $m>0$, only prograde modes are found, while for $m < 0$, only retrograde modes are accessible. As a result, these modes have traditionally been labeled as “regular modes.” To obtain the missing retrograde modes when $m > 0$ or prograde modes when $m < 0$, one usually relies on the symmetry \eqref{eq:MirrorMode}, referring to these solutions as “mirror modes.” Thus, a mode with the overtone index $n$ must acquire yet another label $+/-$ to distinguish between ``regular" and ``mirror", i.e. $\omega_{n \ell m; \pm}$ ---  see e.g.~\cite{MaganaZertuche:2021syq} and references therein.

\smallskip
However, unlike the physically meaningful distinction between prograde and retrograde modes—which relates to the behavior of constant-phase surfaces—the regular/mirror mode terminology is merely a convention arising from past computational strategies. Besides, for $m\neq 0$ it associates $4$ distinct eigenvalues under the same overtone label $n$. Take for instance the quadrupole mode $(\ell, |m|) = (2,2)$. The table in ref.~\cite{BertiRingdown} associates the following values to the mode with $n=0$ for $a/M = 0.8$
\bea
M \omega_{0\, 2\, 2; +} \approx 0.586 - 0.076 \, i,  &\quad&  M \omega_{0\, 2\, -2; +} \approx 0.303 -0.088 \, i \\
M \omega_{0\, 2\, -2; -} \approx -0.586 - 0.076 \, i ,  &\quad&  M \omega_{0\, 2\, 2; -} \approx -0.303 -0.088 \, i.
%M \omega_{0\, 2\, 2; +} \approx 0.464 - 0.086 \, i,  \quad  M \omega_{0\, 2\, -2; -} \approx 0.464 + 0.086 \, i, \\
%M \omega_{0\, 2\, -2; +} \approx 0.324 -0.089 \, i,  \quad  M \omega_{0\, 2\, 2; -} \approx 0.324 + 0.089 \, i. 
\eea
The first line corresponds to the regular modes and the values are directly accessible in the tables of ref.~\cite{BertiRingdown}. The second line correspond to the mirror modes, and they are indirectly derived via the relation \eqref{eq:MirrorMode}. On the other hand, the first column provides the prograde modes, and they are clearly a fundamental modes in the sense of having the slowest decay rate. The second column lists the retrograde modes, which  constitute first overtones in the sense that they have the second slowest decay rate. Hence, it becomes evident that labelling all these modes by $n=0$ is misleading if one wants to convey an intuitive notion of ordering the modes by their decay rate.

\smallskip
Treating the problem as a genuine $m$-mode eigenvalue problem, where $m>0$ and $m<0$ correspond to independent sectors, reduces the need for such distinctions between regular and mirror modes because the modes $\omega_{0\, 2\, 2; +}$ and $\omega_{0\, 2\, 2; -}$ are directly available in the solver as distinct eigenvalues, say with label $q=0$ and $q=1$.

\begin{figure}[h!]
    \centering
    \includegraphics[width=0.49\columnwidth]{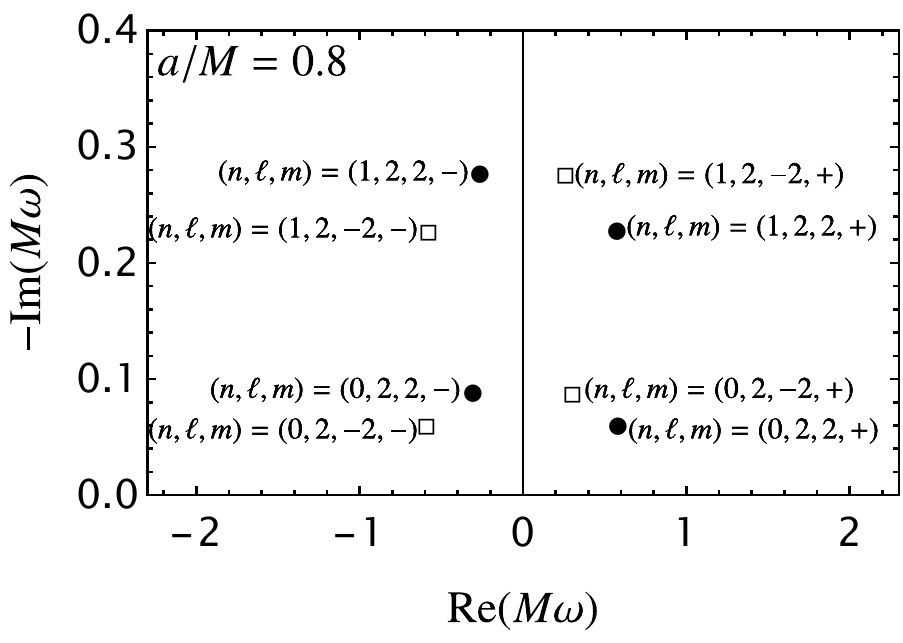}
    \includegraphics[width=0.49\columnwidth]{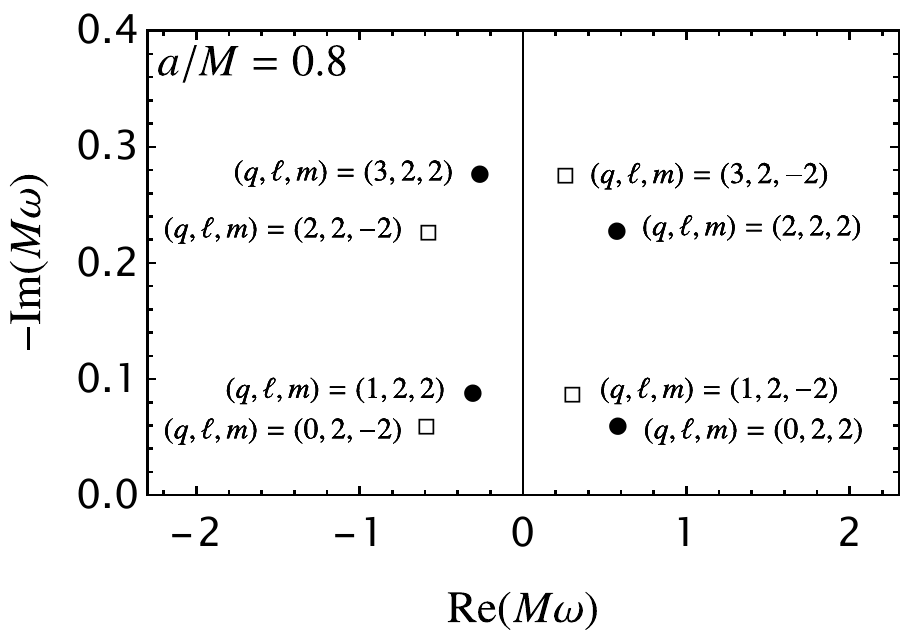}
    \caption{Representation of the QNM modes with notations $(n,\ell,m,\pm)$ on the left panel and $(q,\ell,m)$ on the rigth panel. Dots correspond to QNMs in the $m=2$ sector, while squares to QNMs obtained in the $m=-2$ sector. The traditional notation associates $4$ eigenvalues to each index $n$, which further labels $\pm$ distinguishing ``regular" and ``mirror" modes. Alternative notations associate an {\em unique} index $q$ to each individual overtone in the spectra of a given $m$-sector. Here, even $q=2n$ are prograde modes, whereas odd $q=2n+1$ are retrograde modes, whereas Ref.~\cite{Bourg:2025lpd} have even and odd $q$'s labelling, respectively, regular and mirror modes.}
    \label{berti-qnm}
\end{figure}

Therefore, we suggest a new index for the QNM overtones unifying the notation based on the following physical property. Akin to ref.~\cite{Bourg:2025lpd}, we will employ the triple $(q,\ell,m)$ --- as opposed to $(n\, \ell \,m)$ -- to label a QNM $\omega_{q,\ell, m}$, with the commas in the subscripts emphasising the new notation. Fig.~\ref{berti-qnm} exemplify the $\omega_{n \ell m;\pm}$ notation against the suggested $\omega_{q, \ell, m}$.

For any QNM decay rate $T = 1 / \left| {\rm Im} (\omega) \right|$ and oscillatory frequency $f = \left| {\rm Re} (\omega)\right| /(2\pi)$ we associated a given overtone label $q$. There exists two eigenvalues yielding the specific physical time scales $(T, f)$, but they belong to distinct $m-$sectors, namely QNM $\omega_{q, \ell, |m|}$ and $\omega_{q, \ell, -|m|}$, related to each other via the symmetry condition \eqref{eq:MirrorMode}.

If we adopt the convention that the overtone index $q$ orders QNMs according to their decay rate, i.e.,
\beq
\label{eq:Im_omega_q}
\left|   {\rm Im} (\omega_{0, \ell, m})    \right|  \leq  \left|   {\rm Im}  ( \omega_{1, \ell, m} ) \right| < \left|   {\rm Im}  (\omega_{2, \ell, m}) \right| \leq \left|   {\rm Im}  (\omega_{3,\ell, m})     \right|  < \cdots,
\eeq
then we find that even-indexed modes correspond to prograde solutions, while odd-indexed modes correspond to retrograde solutions. Hence, one obtains a direct relation between the $q$ and $n$ notation by expressing even and odd-indexed modes, respectevely, $q = 2n$ and  $q= 2n +1$, yielding
\bea
\label{eq:Re_omega_q_m+}
{\rm Re}  ( \omega_{2n , \ell, m} ) > 0, \quad {\rm Re} ( \omega_{2n+1,  \ell,  m} ) < 0 \quad  (m  \geq 0), \\
\label{eq:Re_omega_q_m-}
{\rm Re} ( \omega_{2n ,\ell, m} ) < 0, \quad {\rm Re}  ( \omega_{2q+1, \ell,  m} ) > 0 \quad (m  < 0).
\eea
Note that the case $m=0$ shows a degeneracy with $q=0$ and $q=1$ corresponding to the fundamental mode. Thus, ref.~\cite{Bourg:2025lpd} suggest yet another alternative labelling scheme, associating even $q=2n$ and odd $q=2n+1$, respectively, to the regular and mirror modes, as opposed to the prograde and retrograde suggested here. 

\smallskip
In the next section we present convergence tests for the QNM values obtained with the $2D$ eigenvalue solver. This investigation is motivated by the fundamental properties of the radial fixing and Cauchy horizon fixing gauges.

\subsection{Convergence test}

\begin{figure}[t!]
     \centering
     \begin{subfigure}[b]{0.49\textwidth}
         \centering
         \includegraphics[width=\textwidth]{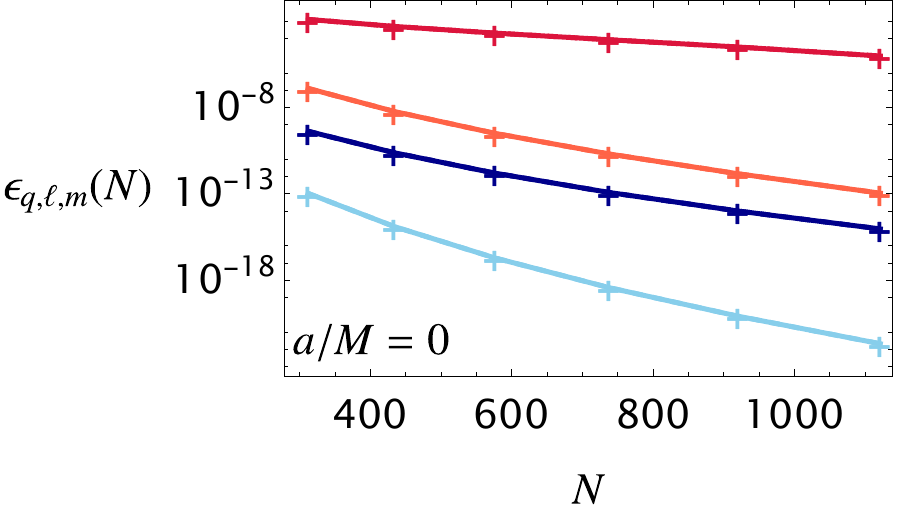}
     \end{subfigure}
     \hfill
     \centering
     \begin{subfigure}[b]{0.49\textwidth}
         \centering
         \includegraphics[width=\textwidth]{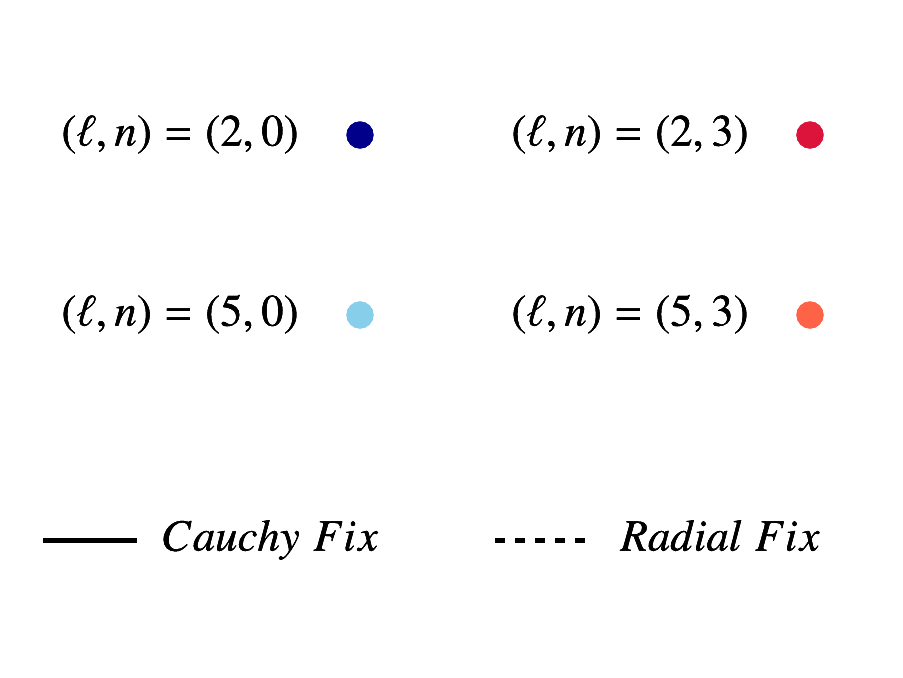}
     \end{subfigure}
     \hfill
     \centering
     \begin{subfigure}[b]{0.49\textwidth}
         \centering
         \includegraphics[width=\textwidth]{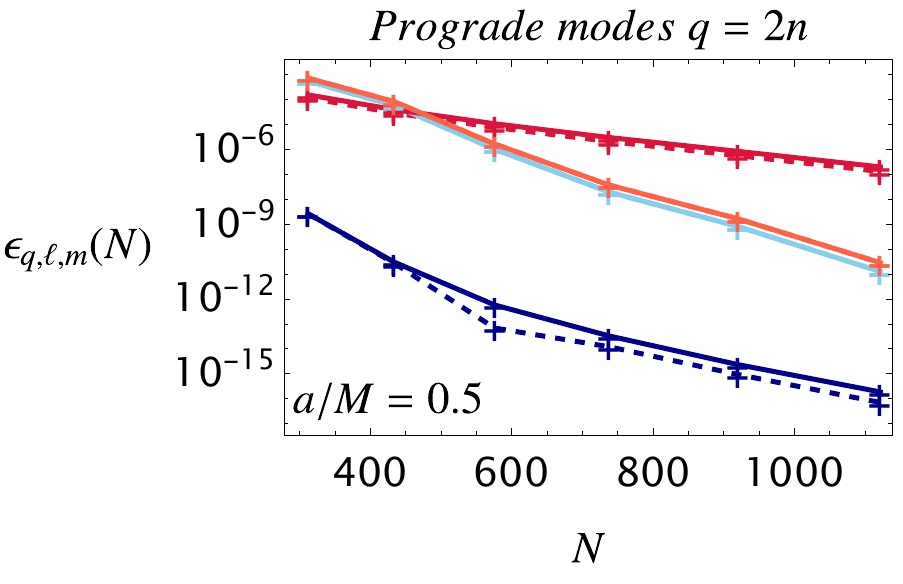}
     \end{subfigure}
     \hfill
     \centering
     \begin{subfigure}[b]{0.49\textwidth}
         \centering
         \includegraphics[width=\textwidth]{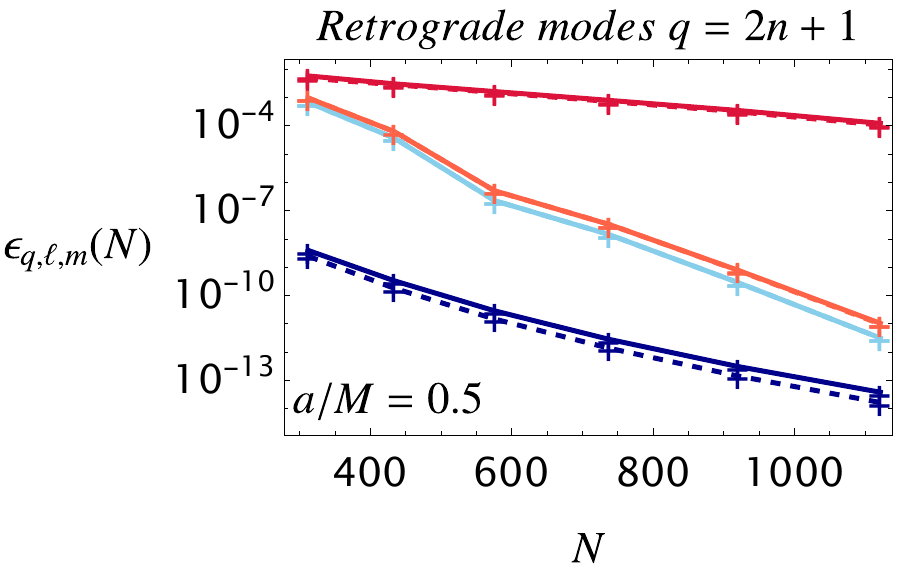}
     \end{subfigure}
     \hfill
     \centering
     \begin{subfigure}[b]{0.49\textwidth}
         \centering
         \includegraphics[width=\textwidth]{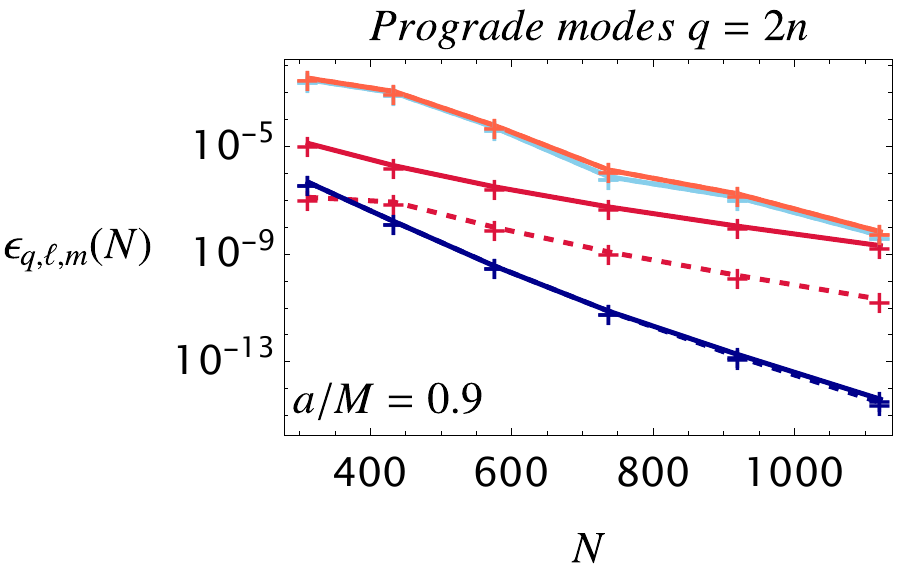}
     \end{subfigure}
     \hfill
     \centering
     \begin{subfigure}[b]{0.49\textwidth}
         \centering
         \includegraphics[width=\textwidth]{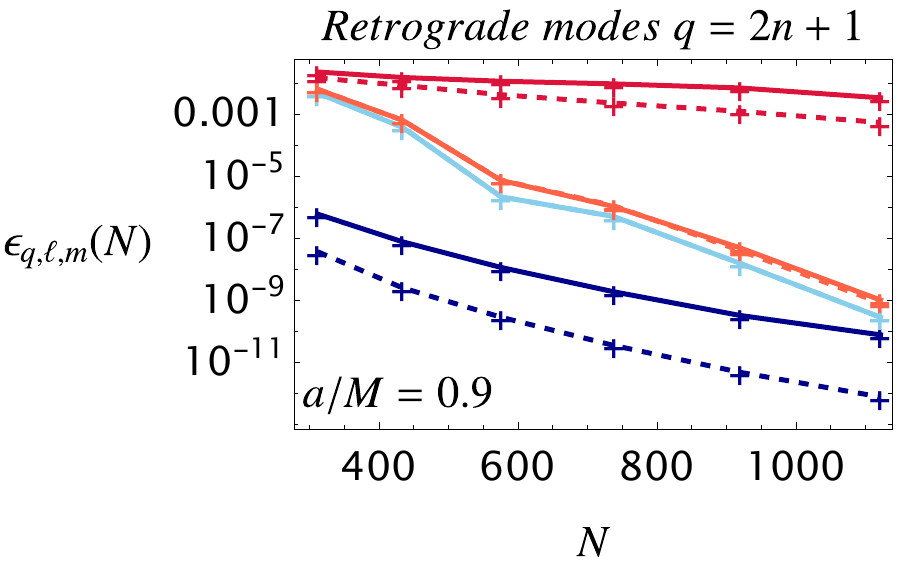}
     \end{subfigure}
     \hfill
    \caption{Relative error $\epsilon_{q,\ell,m}$ as a function of discretized operador $\hat{\boldsymbol L}$'s total size, cf.~\eqref{eq:disc_L}. Convergence tests performed in the $m=2$-sector for both the prograde and retrograde modes, and different individual eigenvalues. Both choices of hyperboloidal foliation --- radial fixing and Cauchy horizon fixing gauges --- show the same performance in terms of accuracy and convergence rate, despite their distinct geometrical properties in the interior black hole region.}
    \label{CVtest}
\end{figure}

\smallskip
As described in Sec.~\ref{sec:BL_to_Hyp}, the radial fixing gauge allows for a smooth transition into the extremal Kerr limit. This property arises from the coordinate location of the Cauchy horizon $\sigma_{c} = \kappa^{-2}$ being well defined for $\kappa \in[0,1]$, with $\sigma_{c} = \sigma_{h} =1$ in the extremal limit. In contrast, in the Cauchy horizon fixing gauge, we have $\sigma_c \rightarrow \infty$, leading to an ill-defined limit as $\kappa \rightarrow 0$, with the QNM operator scaling as $(1-\kappa)^{-1}$.

\smallskip
While this property might suggest an advantage for the radial fixing gauge, a QNM solver based on the Cauchy horizon fixing gauge has an exact correspondence with Leaver’s algorithm. With this choice, a Taylor expansion of the QNM eigenfunction around the horizon remains valid for all $\kappa \in [0,1)$. This property is not valid in the radial fixing gauge, since Taylor expansions that converge around the horizon and extend to future null infinity ($\scri^+$ at $\sigma =0$) remain valid only while the Cauchy horizon $\sigma_{c}$ lies outside the convergence radius, i.e., for $\kappa^2 < 1/2$. 

\smallskip
Given these different properties between both coordinate choices, a systematic study of the solver's accuracy and efficiency for the available hyperboloidal parametrisations is necessary. For this purpose, we parametrize the truncation parameters in Eqs.~\eqref{eq:nodes_1} and \eqref{eq:nodes_2} as $N_1 = 5 N_2 = N$ and define the relative error as
\beq
\label{relativeerrorNtot}
\epsilon_{q, \ell, m}(N) = \dfrac{\vert \omega_{q, \ell, m}(N) - \omega_{q, \ell, m}^{{\rm Ref}} \vert}{\vert \omega_{q, \ell, m}^{{\rm Ref}} \vert},
\eeq
where $\omega_{q, \ell, m}^{\rm Ref}$ denotes the QNM values obtained with a high-resolution reference calculation using $N_{\rm ref}=55$. Our objective is to determine whether one hyperboloidal foliation offers greater numerical efficiency than the other. 

\smallskip
Fig.~\ref{CVtest} display the convergence results within the sectors $m=2$ for prograde ($q=2n$) and retrograde ($q=2 n+1$). We select the representative examples with $(\ell,n) = (2,0)$ (dark blue) and $(\ell,n) = (5,0)$ (light blue), and their corresponding overtones $(\ell,n) = (2,3)$ (red) and $(\ell,n) = (5,3)$ (orange). The different panels display the results for several BH parameter: $a/M = 0$ (top), $a/M = 0.5$ (middle), $a/M = 0.9$ (bottom), with results for the radial fixing gauge being displayed as dotted lines and the Cauchy horizon fixing gauge in solid lines. 

\smallskip
We do not find any distinction between the two gauges, with the convergence rate for the QNMs being the same for both options in all the studied cases. We observe however, rather slow convergence rates for retrograde modes as they approaches the imaginary axis for increasing values of  the BH rotation parameter. This property is a direct consequence of the presence of spurious modes clustering near the branch cut as $|\kappa| \lesssim 1$.

\smallskip
Even though we do not identify any difference in the convergence rate for the numerical errors associates with the radial or Cauchy fixing gauges, these options do offer different representations for the underlying QNM eigenfunctions.

\subsection{QNM Eigenfunctions}
Having benchmarked the QNM spectrum as eigenvalues of the system, we now study the corresponding QNM eigenfunctions, denoted by
\beq
\spinHyperPhiQNM{\spin}{q}{\ell}{m}(\sigma,x) := \spinHyperPhioQNM{\spin}(\sigma,x)
\eeq

\begin{figure}[h!]
    \centering
    \includegraphics[width=0.42\columnwidth]{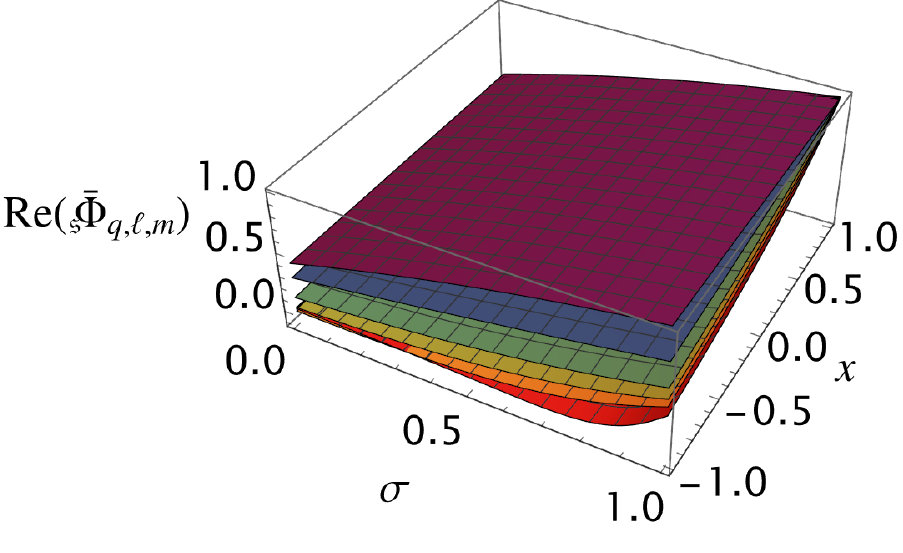}
    \includegraphics[width=0.57\columnwidth]{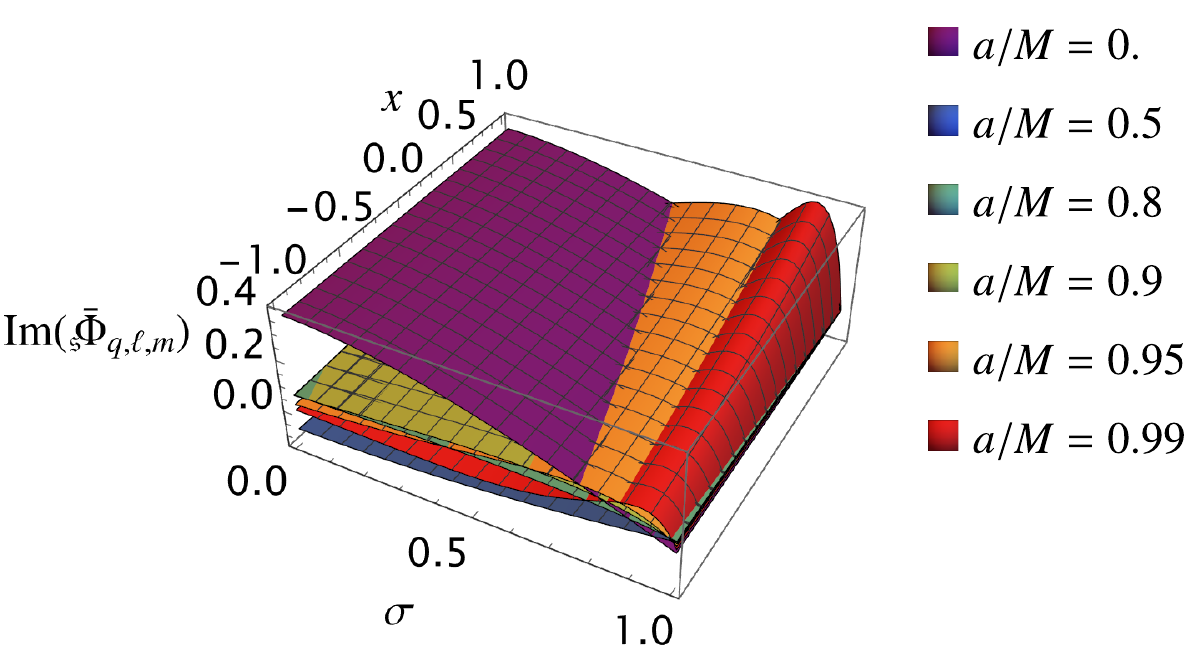}
    \caption{ Eigenfunctions associated with the mode $(q,\ell,m) = (0,2,2)$ computed using the radial fixing gauge. On the left panel the real part of the function is represented while on the right panel the imaginary part is. The hyperboloidal framework allows for a regular representation of the the QNM eigenfuctions in the entire domain $(\sigma,x)\in[0,1]\times[-1,1]$.}
    \label{fig:QNN_Eigenfuncitons}
\end{figure}

Since the eigenfunctions $\spinHyperPhiQNM{\spin}{q}{\ell}{m}(\sigma, x)$ satisfy the homogenous equation ~\eqref{eq:HypTeuk_PDE} they are uniquely define up to an overall multiplication constant. We fix this degree of freedom by normalising the eigenfunctions to 
\beq
\spinHyperPhiQNM{\spin}{q}{\ell}{m}(1,1) = 1.
\eeq
This choice agrees with standards from an $\ell m$-decomposition. For instance, typically Leaver's method sets the radial function to unit at the horizon $\sigma=1$, whereas the angular function (e.g. Legendre Polynomials) are also commonly normalised to unit at $x=1$. Fig.~\ref{fig:QNN_Eigenfuncitons} shows the functions resulting from the eigenvalue problem for a several values of the black-hole rotation parameter $a/M$ for the fundamental quadrupole mode $(q,l,m) = (0,2,2)$. As expected from the hyperboloidal formalism, $\spinHyperPhiQNM{\spin}{q}{\ell}{m}(\sigma, x)$ is regular in the entire domain $[\sigma,x]\in [0,1]\times[-1,1]$.

\smallskip
From the common approach based on an $(\ell,m)-$decomposition, there are two property of interest in $2D$ surface $\spinHyperPhiQNM{\spin}{q}{\ell}{m}(\sigma, x)$ concerning the functions behaviour along the lines $x=$constant or $\sigma=$constant. We address these points in the next sections.

\subsubsection{Near extremality and steep gradients\\}
\begin{figure}[h!]
    \centering
    \includegraphics[width=0.49\columnwidth]{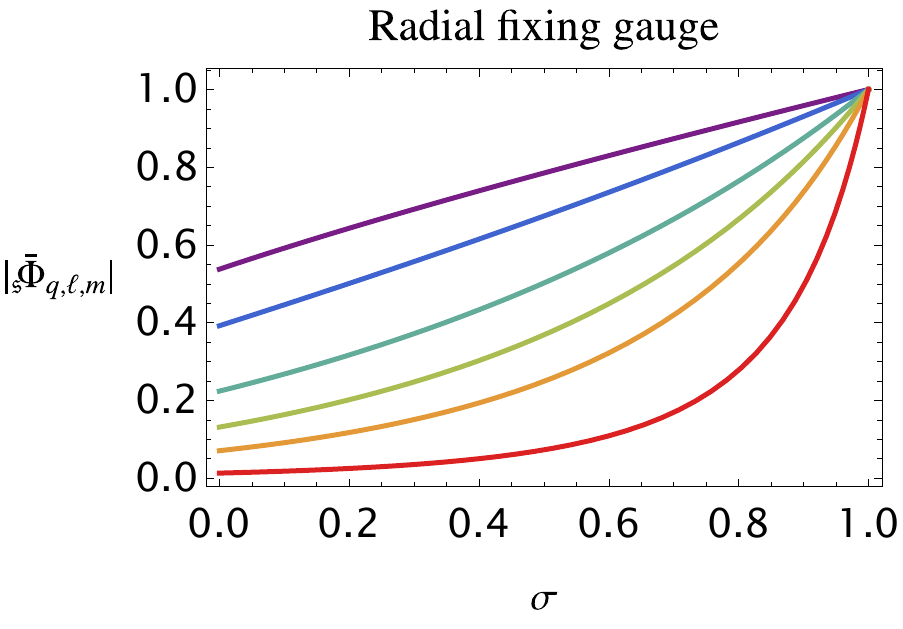}
    \includegraphics[width=0.49\columnwidth]{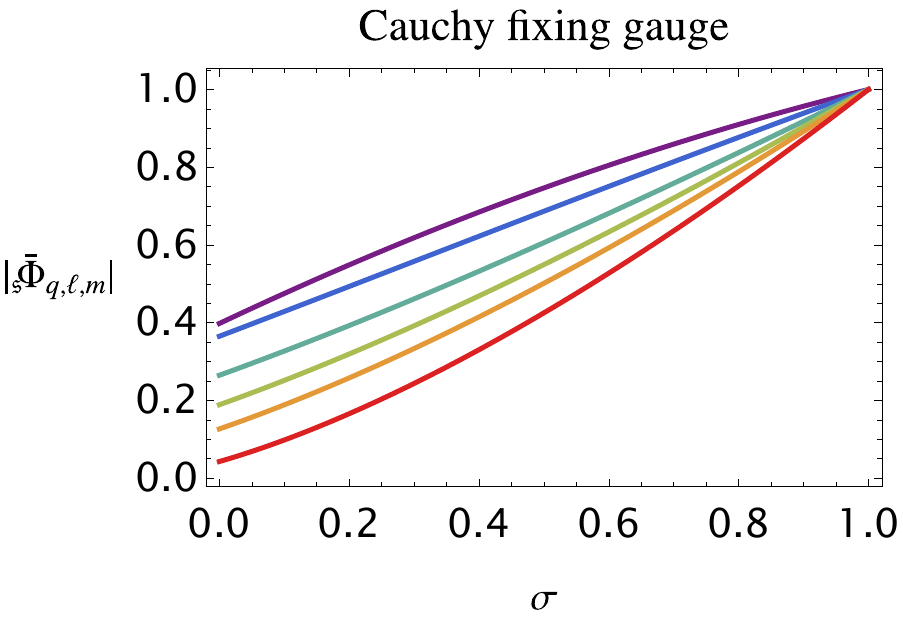}
    \caption{Real part of the eigenfunctions associated with the mode $(n,\ell,m) = (0,2,2)$ projected on the plan $x=1$. Color code used for the legend is given on figure Fig.\ref{fig:QNN_Eigenfuncitons}. {\em Left Panel:} In the radial fixing gauge, the radial functions develop strong gradients around the horizon $\sigma=1$ as the spacetime approaches extremality. {\em Right Panel:} one does not observe such gradients in the Cauchy horizon fixing gauge, indicting that they are consequence of choice for the foliation, and not attached to any physical property of extremal black holes.}
    \label{fig:QNM_radial_Eigenfunctions}
\end{figure}

Firstly, we consider the radial behaviour along a given $x=$constant. When formulating the hyperboloidal QNM problem for the angular $+$ radial ODES, Ref.~\cite{Ripley:2022ypi} observed the radial eigenfunctions developing strong gradients around the horizon ($\sigma =1$) as the Kerr spacetime approached extremality, and as mentioned in sec.~\ref{sec:Kerr}, Ref.~\cite{Ripley:2022ypi} works in the radial fixing minimal gauge. The left panel of fig.~\ref{fig:QNM_radial_Eigenfunctions} displays the corresponding eigenfunction $\spinHyperPhiQNM{\spin}{q}{\ell}{m}(\sigma, x)$ for the fundamental quadrupole mode $(q,l,m) = (0,2,2)$ along the axis $x=1$, where we also observe the gradients developing around $\sigma=1$ as $a/M \to 1$.

\smallskip
One hypothesis raised is that such gradients might be related to the Aretakis instability\cite{Aretakis:2011gz,Aretakis:2012ei}, prompting the need for further studies in the near extremal limit. Here, however, we address the issue from a different perspective, enquiring whether the behaviour around $\sigma=1$ is a coordinate effect, resulting for the particular choice of hyperboloidal system. Hence, fig.~\ref{fig:QNM_radial_Eigenfunctions}'s right panel shows the corresponding $\spinHyperPhiQNM{\spin}{q}{\ell}{m}(\sigma, 1)$ resulting from the Cauchy horizon fixing gauge. Contrary to the previous results, we do not observe any strong gradients developing around $\sigma =1$, indicating that the gradients are a consequence of the particular hyperboloidal foliation.

\subsubsection{$\ell m-$projection into harmonics\\}
\begin{figure}[t!]
    \centering
    \includegraphics[width=0.47\columnwidth]{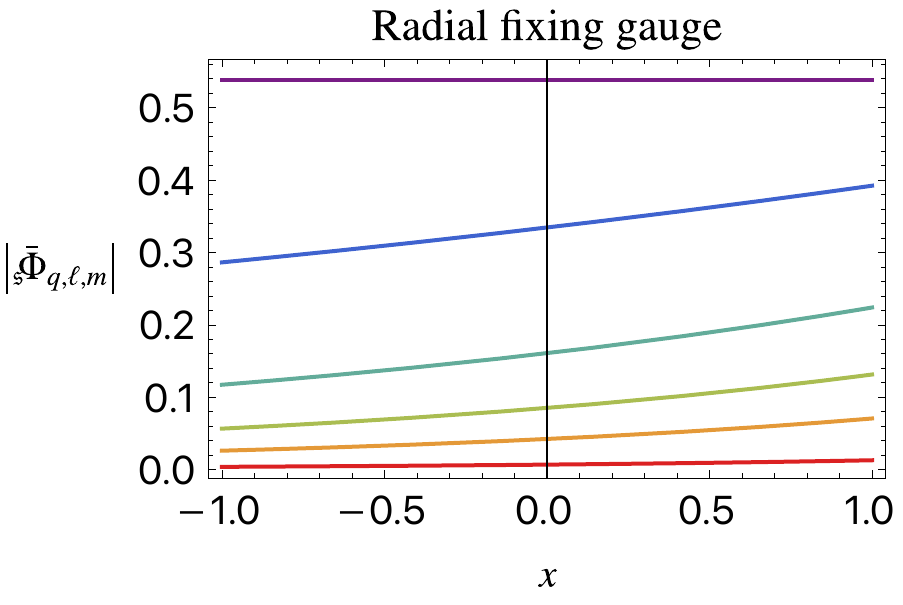}
        \includegraphics[width=0.52\columnwidth]{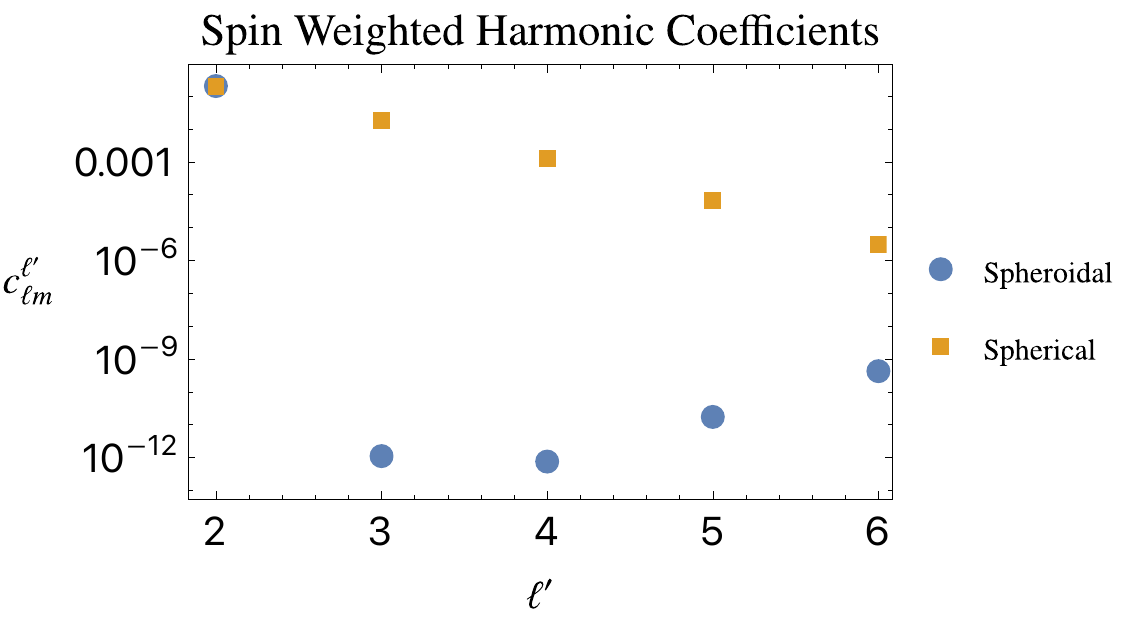}
    \caption{Angular structure of QNM eigenfunctions $\spinHyperPhiQNM{\spin}{q}{\ell}{m}$ for $(q,\ell,m) = (0,2,2)$ evaluated at future null infinity ($\sigma = 0$). Color code used for the legend is given on figure Fig.\ref{fig:QNN_Eigenfuncitons}. {\em Left panel:} The mild variation with respect to the angular coordinate $x \in [-1,1]$ is a consequence of regularisation the behaviour at the symmetry axis, cf.~\eqref{eq:TeukFunc_BL_HypRadial}. {\em Right panel}: Harmonic decomposition coefficients $c^{\ell’}_{\ell m}$ obtained by projecting the angular profile onto spin-weighted spheroidal harmonics ${}_\spin S_{\ell’ m}(x; a\omega)$ (blue circles) and spherical harmonics ${}_\spin S_{\ell’ m}(x)$ (orange squares). In the spheroidal case, the expected orthogonalityt $c^{\ell’}_{\ell m} \propto \delta^{\ell'}_\ell$ confirms the alignment of the solution with the natural basis for the angular. In contrast, the spherical harmonic decomposition shows the expected spectral convergence. This flexibility of the $m$-mode approach highlights its utility in GW applications requiring projections in either basis.}
    \label{fig:QNM_angularl_Eigenfunctions}
\end{figure}

Secondly, we analyse the projection of the eigenfunctions at fixed values of $\sigma$. The left panel of Fig.\ref{fig:QNM_angularl_Eigenfunctions} shows the angular profile of the QNM eigenfunction at future null infinity, $\sigma = 0$. Although these functions are inherently constructed from harmonic components, we observe a relatively mild variation with respect to the angular coordinate $x \in [-1,1]$. This behaviour arises directly from the definition of the conformal master function $\spinHyperPsi{\spin}$ in eq.\eqref{eq:TeukFunc_BL_HypRadial}, which explicitly factors out the dominant angular behaviour $\sim (1+x)^{\delta_1/2}(1-x)^{\delta_2/2}$, required to regularise the function at the symmetry axis $x = \pm 1$.

\smallskip
We recall that the PDE \eqref{eq:HypTeuk_PDE}, formulated through an $m$-mode decomposition, has no {\em a priori} knowledge of a preferred harmonic basis for the angular functions. In principle, one may decompose the angular dependence using any spin-weighted basis functions ${}_\spin B_{\ell m}(x)$. More specifically, a $(\ell,m)$-decomposition of the $2D$ solution $\spinHyperPhio{\spin}(\sigma, x)$ takes the form
\beq
(1+x)^{\delta_1/2}(1-x)^{\delta_2/2} \, \spinHyperPhiQNM{\spin}{q}{\ell}{m}(\sigma, x) = \sum_{\ell’ = |\spin|}^{\infty} c^{\ell’}_{\ell m} \, {}_\spin B_{\ell’ m}(x),
\eeq
with the coefficients $c^{\ell’}_{\ell m}$ given by standard projection integrals:
\beq
c^{\ell’}_{\ell m} = \int_{-1}^1 dx \,  (1+x)^{\delta_1/2}(1-x)^{\delta_2/2} \, \spinHyperPhiQNM{\spin}{q}{\ell}{m}(\sigma, x) \, {}_\spin B_{\ell’ m}(x).
\eeq

\smallskip
Importantly, we are not restricted to a specific choice of basis functions. This flexibility is particularly useful in the context of gravitational-wave astronomy. Although the natural basis arising in the $(\ell,m)$ decomposition of the Teukolsky equation is the spin-weighted {\em spheroidal} harmonics ${}_\spin S_{\ell’ m}(x; a\omega)$, the {\em spherical} harmonic basis ${}_\spin S_{\ell’ m}(x)$ often provides a more consistent and broadly applicable representation for the angular dependence. In fact, spheroidal harmonics are frequently re-expanded in terms of spherical harmonics to facilitate comparison with observational templates.

\smallskip
The $m$-mode approach enables one to freely choose the desired basis. This flexibility is illustrated in the right panel of Fig.~\ref{fig:QNM_angularl_Eigenfunctions}, which shows the harmonic coefficients $c^{\ell’}_{\ell m}$ computed using both {\em spheroidal} harmonics ${}_\spin S_{\ell’ m}(x; a\omega_{q,\ell,m})$ and {\em spherical} harmonics ${}_\spin S_{\ell’ m}(x)$. In the former case (blue circles), the expected result $c^{\ell’}_{\ell m} \propto \delta^{\ell’}_\ell$ is recovered (up to numerical roundoff error), confirming the eigenfunction’s spheroidal character. In the latter case (orange squares), we observe the expected exponential decay in the expansion coefficients, reflecting the spectral decomposition of a spheroidal harmonic into spherical harmonics.

\section{Conclusion}\label{sec:Conclusion}
In this work, we have revisited the problem of calculating QNM in the Kerr spacetime by adopting an alternative treatment of the Teukolsky equation. The strategy combines two emergent approaches in black hole perturbation theory: (i) a \emph{hyperboloidal formulation}, resolving well-known limitations in the representation of QNM eigenfunctions near the asymptotic regions $r \rightarrow \rh$ and $r \rightarrow \infty$; and (ii) an $m$-mode decomposition of the Teukolsky equation, expressing it as a $2D$ PDE in a domain defined by radial and angular directions. For a fixed $m$-value, such an approach allows the formulation of the QNM problem as a genuine single eigenvalue problem, enabling the simultaneous calculation of multiple modes—traditionally labelled by the indices $n$ and $\ell$—which would otherwise be computed individually with solvers requiring a well guessed initial seed feeding a root-finder algorithm.

\smallskip

Even though the Kerr QNM values are well-documented in the literature, their classification remains intricate. This complexity arises from the rich structure of the Kerr QNM spectrum, where black hole rotation breaks the degeneracy in the QNM values present in the Schwarzschild case within each $m$-mode sector. It is common in the literature to find tables where the overtone index $n$ labels modes that are continuously deformed from their Schwarzschild counterparts. As a result, such tables often assign the same index $n$ to up to four distinct modes—typically including both prograde and retrograde branches with differing decay rates and oscillation frequencies. This labelling obscures key physical distinctions between the modes. By contrast, the QNM spectra obtained through the $m$-mode decomposition strategy enable a more transparent classification: a single index $q$ uniquely identifies each overtone as a point in the spectra, avoiding the conventional—but ultimately artificial—distinction between “regular” and “mirror” modes.

\smallskip
Apart from reproducing the Kerr QNM values, we performed convergence tests to investigate whether different choices of hyperboloidal foliations—each with distinct geometric treatments of the Cauchy horizon—affect the efficiency of the solver. Recall that in the radial fixing gauge, the Cauchy horizon is smoothly foliated by hyperboloidal surfaces, with its coordinate location depending explicitly on the black hole rotation parameter $\kappa = a/\rh$. In this formulation, the extremal limit is recovered smoothly as $\left| a/M \right| \to 1$. In contrast, the Cauchy horizon fixing gauge locates the Cauchy horizon at a fixed coordinate value $\sigma_{c} \to \infty$, meaning the hyperboloidal slices do not foliate it, and the extremal limit becomes discontinuous, transitioning instead to the near-horizon geometry. Our convergence tests demonstrate that both the radial fixing and Cauchy horizon fixing gauges yield QNM values with comparable numerical precision, exhibiting the same exponential convergence rates. This suggests that, despite their differing geometric constructions—especially regarding the extremal limit—both gauges provide robust and equally efficient frameworks for solving the QNM eigenvalue problem within the $2D$ spectral approach.

\smallskip
The choice of foliation, however, does have a notable impact on the radial behaviour of the eigenfunctions near the event horizon, particularly in the extremal Kerr regime. Previous studies~\cite{Ripley:2022ypi} have reported strong gradients forming near the horizon in this limit, sometimes interpreted as physical features of extremal black holes. Our results show that these gradients are, in fact, coordinate artefacts: they arise from the specific hyperboloidal slicing used rather than any intrinsic geometric or physical property of the spacetime.

\smallskip
In addition, we also examined the angular structure of the corresponding eigenfunctions. A key advantage of the $m$-mode formulation is its flexibility in representing the angular dependence: since the angular profile is computed directly as part of the $2D$ solution, one may project it onto any desired spin-weighted harmonic basis. We demonstrated this by decomposing the eigenfunctions into both spin-weighted spheroidal and spherical harmonics. In the former case, the coefficients exhibit the expected orthogonality property, confirming the solution’s alignment with the standard $(\ell,m)$ separation ansatz. In the latter, the projection reveals the characteristic exponential decay of spheroidal harmonics represented in a spherical basis. These results highlight the adaptability of the $m$-mode approach and its utility for interfacing with gravitational waveform models, which often favour spherical harmonic decompositions.

\smallskip
This work also lays the foundation for several future directions. While root-finding algorithms such as Leaver’s continued fraction method remain highly efficient when good initial guesses are available, our $m$-mode solver removes the need for any prior knowledge of the spectrum. It delivers the entire set of eigenvalues in a single computation, making it a powerful tool for constructing accurate seeds for refinement by other methods. Most importantly, the infrastructure is the gateway for systematic treatment and understanding of the non-normal nature of the Kerr QNM operator. It shall be employed as the primary tool for identifying QNM instabilities, as the full QNM eigensystem obtained from the $m$-mode decomposition naturally provides the basis for expanding pseudospectrum analysis in axisymmetric spacetimes\cite{Cai:2025irl}. In this context, the algorithm developed here is directly applicable to the study of QNM excitation amplitudes and late-time tails through a modal expansion approach. In particular, it offers the framework for implementing the promising Keldysh expansion, where mode projections can be computed using the properties of the numerically obtained transposed operator~\cite{Besson:2024adi}.

\section*{Acknowledgments}
The Tycho supercomputer hosted at the SCIENCE HPC center at the University of Copenhagen was used for supporting this work.
JA thanks the Strong Group at the Niels Bohr Institute (NBI) for their kind hospitality during the initial stages of this work.
RPM acknowledges support from the Villum Investigator program supported by the VILLUM Foundation (grant no. VIL37766) and the DNRF Chair program (grant no. DNRF162) by the Danish National Research Foundation. 
The Center of Gravity is a Center of Excellence funded by the Danish National Research Foundation under grant No. 184.

\section*{Bibliography}
\bibliographystyle{unsrt.bst}
\bibliography{bibitems}

\end{document}